\newcounter{myctr}
\def\myitem{\refstepcounter{myctr}\bibfont\noindent\ifnum\themyctr>9\else\phantom{0}\fi\hangindent17pt\themyctr.\enskip}

\documentclass{ws-ijqi}
\usepackage{hyperref}
\usepackage[super,sort,compress]{cite}

\usepackage{braket}
\usepackage[caption = false]{subfig}%
\begin{document}

\catchline{}{}{}{}{}

\title{QUANTUM ARITHMETIC OPERATIONS BASED ON QUANTUM FOURIER TRANSFORM ON SIGNED INTEGERS}

\author{Engin \c{S}ahin}

\address{Department of Computer and Instructional Technologies Education,\\ \c{C}anakkale Onsekiz Mart University, \\ \c{C}anakkale 17100 Turkey\\
enginsahin@gmail.com}

\maketitle

\begin{history}
\received{Day Month Year}
\revised{Day Month Year}
\end{history}

\begin{abstract}
The quantum Fourier transform brings efficiency in many respects, especially usage of resource, for most operations on quantum computers. In this study, the existing QFT-based and non-QFT-based quantum arithmetic operations are examined. The capabilities of QFT-based addition and multiplication are improved with some modifications. The proposed operations are compared with the nearest quantum arithmetic operations. Furthermore, novel QFT-based subtraction, division and exponentiation operations are presented. The proposed arithmetic operations can perform non-modular operations on all signed numbers without any limitation by using less resources. In addition, novel quantum circuits of two's complement, absolute value and comparison operations are also presented by using the proposed QFT based addition and subtraction operations.
\end{abstract}



\markboth{Engin \c{S}ahin}
{Quantum arithmetic operations based on quantum Fourier transform on signed integers}

\section{Introduction}
\label{intro}
In recent years, quantum computation have started to attract attention with the ease of solving difficult problems in classical computation. After Shor proposed the quantum factoring algorithm \cite{Shor1994}, researchers' interest in quantum arithmetic operations increased. Quantum arithmetic operations are required in many studies such as quantum signal processing, quantum machine learning. Especially in quantum image processing, arithmetic operations are used in many processes such as steganography, edge detection and pattern recognition.%

The modular adder, the modular multiplier and the modular exponentiation operations proposed by Vedral et al. \cite{Vedral1996} are the first elementary quantum arithmetic operations to improve the time and the memory complexity of Shor's quantum factoring algorithm. Gossett \cite{Gossett1998} showed how to design modular arithmetic elements from quantum gates using the carry-save method on a classic computer. Draper \cite{Draper2000} proposed a new method for computing sums on a quantum computer. This technique uses the quantum Fourier transform (QFT) and reduces the number of carry qubits. Cuccaro et al. \cite{Cuccaro2004} presented a new linear-depth ripple-carry quantum addition circuit. This new adder circuit uses only a single ancillary qubit instead of the many ancillary qubits used by the previous adder circuits. Takahashi and Kunihiro \cite{Takahashi2005} proposed a quantum circuit based on the ripple-carry approach for addition of two $n$-bit binary numbers that uses no ancillary qubits. Draper et al. \cite{Draper2006} focused on decreasing the depth and constructed a fast quantum circuit for addition using the classical carry-lookahead technique. Takahashi and Kunihiro \cite{Takahashi2008} combined a modified version of Draper et al.’s quantum carry-lookahead adder in Ref.~\citen{Draper2006} with parallel applications of Takahashi et al.’s quantum ripple-carry adder in Ref.~\citen{Takahashi2005}. This modified adder used a few qubits and saved ancillary qubits. Alvarez-Sanchez et al. \cite{Sanchez2008} proposed a quantum multiplication architecture based on the Booth algorithm on classical computers, which used signed integers. Markov and Saeedi \cite{Markov2012} proposed constant-optimized quantum circuits for modular multiplication and exponentiation. Wang et al. \cite{Wang2016} presented an improved linear-depth ripple-carry quantum addition circuit, which is an elementary circuit used for quantum computation. Compared with previous adder circuits costing at least two Toffoli gates for each bit of output, the proposed adder used only a single Toffoli gate. Babu \cite{Babu2017} proposed a cost-efficient quantum multiplier–accumulator unit. Babu presented a fast multiplication algorithm with the optimum time complexity and designed a novel quantum multiplier based on the proposed algorithm. Ruiz-Perez and Garcia-Escartin \cite{Perez2017} made some simple variations to the existing QFT adders and multipliers, proposed a novel adder and a novel multiplier based on QFT with improved capabilities. The modified circuits could perform non-modular addition and modular multiplication operations. These circuits could also perform modular signed addition for numbers up to $d/2$ ($d$ is the number of modulo).%

The overall goal of quantum arithmetic operations should be to be able to operate on all numbers without any limitation, using as few ancillary qubits and basic quantum gates as possible. All the arithmetic operations in the literature are on integers with same numbers of qubit $n \times n$. All the studies except the adder in Ref.~\citen{Perez2017} are modular arithmetic operations. In addition, only the QFT adder in Ref.~\citen{Perez2017} and the multiplier in Ref.~\citen{Sanchez2008} can operate with signed integers. Furthermore, too many $n$-CNOT gates are used in non-QFT based arithmetic operations. $n$-CNOT gate consists of $2n-1$ Toffoli gates (each Toffoli gate consists six CNOT gate) and one CNOT gate. This is also leading to increase in time complexity. Most of the studies in the literature have focused on addition, multiplication and exponentiation to make Shor's quantum factoring algorithm in Ref.~\citen{Shor1994} more efficiently.%

In this paper, both modular and non-modular QFT based addition, subtraction, multiplication, division and exponentiation arithmetic operations are presented. In addition to these arithmetic operations, two's complement, absolute value calculation and comparison operations are presented. All the arithmetic operations are performed on two signed integers with the different number of qubits ($n \times m$). Hence, the proposed operators will be the first in the literature. The comparison operators in the literature compare only positive integers with the same number of qubits. The proposed comparison operation also compares signed integers (positive and negative) with the different number of qubits. The results of multiplication, division and exponentiation operations are calculated on a separate qubit sequence, not on input qubits. Thus, the required resource and the time complexity in operations are reduced by reusing the input qubits. The proposed quantum arithmetic operators use less ancillary qubits and basic quantum gates on all integers with no limitations in accordance with the overall goal. Moreover, since the proposed operators are QFT-based, they are more efficient than conventional calculation based operators.%

This paper is organized as follows. Section~\ref{sec:2} briefly introduces the two's complement of signed binary integers and the quantum Fourier transform. Section~\ref{sec:3} presents the quantum circuits for addition, subtraction, multiplication, division, exponentiation, two's complement, absolute value and comparison operations and briefly explanations of the related operations. Section~\ref{sec:4} makes comparisons between the proposed operations and the relevant operations with respect to the usage of ancillary qubits and time complexity. A conclusion remarks are given in Sect~\ref{sec:5}.

\section{Preliminaries}
\label{sec:2}

\subsection{Two's complement of signed binary integers}
\label{sec:2.1}

Two's complement is the most common method of representing signed integers on computers. The two's complement is calculated by inverting the bits and adding one. In this scheme, if the binary number $(010)_2$ encodes the signed integer $(2)_{10}$, then its two's complement, $(110)_2$, encodes the inverse $(-2)_{10}$. The leading bit is the sign bit and is called the most significant bit (MSB). If the MSB of a number is 0, the number is positive, and if the MSB of a number is 1, the number is negative. Any $n$-bit integer $x=x_0x_1 \cdots x_{n-1}$ can be represented as a quantum state $\ket{x}$ as follows:

\begin{equation}
\ket{x} =\ket{x_0} \otimes \ket{x_1} \otimes \cdots \otimes \ket{x_{n-1}} = \ket{x_0x_1 \cdots x_{n-1}}. \label{eq1}%
\end{equation}

where $x_i \in \{0, 1\}$, $i=0, 1, \cdots , n-1$ and $\otimes$ is a tensor product. If $x$ is positive, then $x_0=0$ and $x$ is simply represented as a binary sequence. In the other case, if $x$ is negative, then $x_0=1$ and $x$ is represented by the two's complement of signed integer as follows.

\begin{eqnarray} 
&X^{\otimes n}\ket{x} = \ket{\bar{x}}= \ket{\bar{x}_0 \bar{x}_1 \cdots \bar{x}_{n-1}}. \label{eq2}\\
&\ket{\bar{x}}+\ket{1} = \ket{-x}. \label{eq3}
\end{eqnarray}

Where $X$ is a Pauli X gate. The quantum circuit of the two's complement method using by the proposed QFT addition is given in Sec.~\ref{sec:3.3}.%

\subsection{The quantum Fourier transform}
\label{sec:2.2}

Quantum Fourier transform (QFT) is a application of classical discrete Fourier transform to the quantum states \cite{Nielsen2010,Sahin2018a}. The quantum Fourier transform of state $\ket{x}$ from the computational basis $\ket{0}, \ket{1}, \cdots, \ket{n-1}$ can be defined as follows:%

\begin{equation}
QFT \ket{x} = \frac{1}{\sqrt{n}} \sum_{y=0}^{n-1} e^{i\frac{2 \pi xy}{n}}\ket{y}. \label{Eq:4}%
\end{equation}

The QFT encodes a number $x$ into the phases $e^{i\frac{2 \pi xy}{n}}=\omega^{xy}$ of all states $\ket{y}$ of equal amplitude $1/\sqrt{n}$ in superposition. The state $QFT\ket{x}$ is usually shown as $\ket{\phi(x)}$. The inverse QFT ($QFT^{\dagger}$) can be defined as follows.

\begin{equation}
QFT^{\dagger}\ket{y} = \frac{1}{\sqrt{n}} \sum_{x=0}^{n-1} e^{-i\frac{2 \pi xy}{n}}\ket{x}. \label{Eq:5}%
\end{equation}

\begin{figure}%
	\centerline{
	\subfloat[]{\includegraphics[width=.70\textwidth]{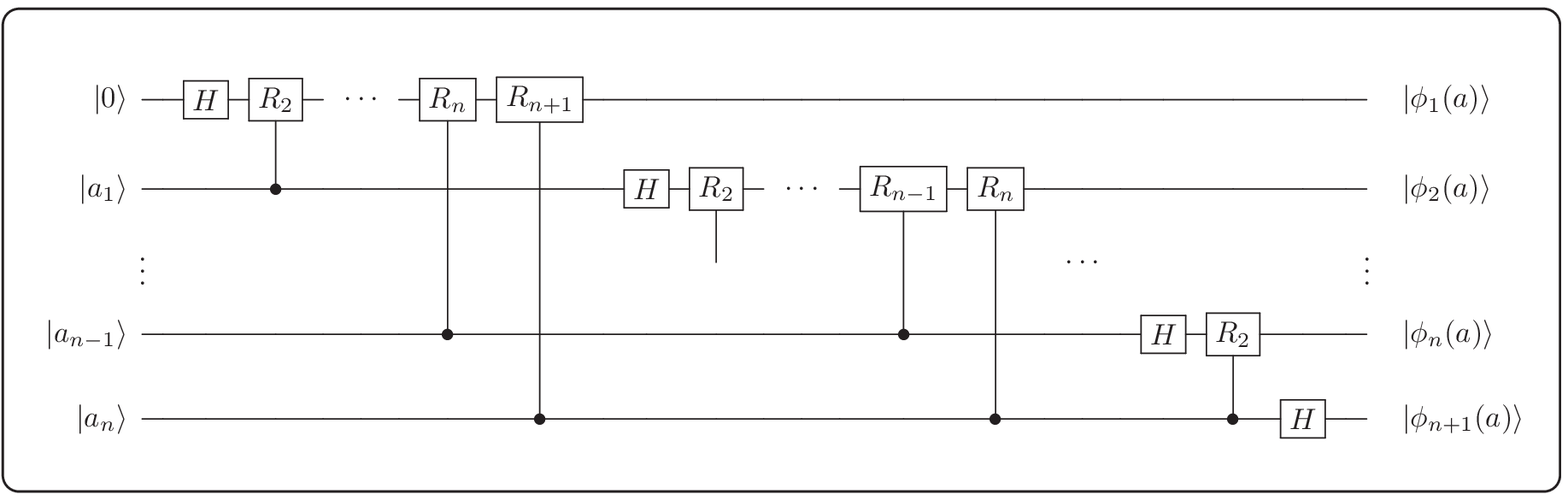}}%
	\hspace{1mm}%
	\subfloat[]{\includegraphics[width=.27\textwidth]{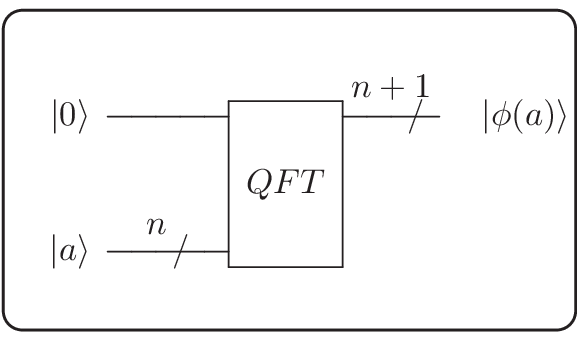}}}\vspace*{8pt}%
	\caption{The quantum Fourier transform (a) the quantum circuit of QFT, (b) the simplified graph of QFT.}%
	\label{Fig:1}%
\end{figure}

\begin{figure}%
	\centerline{
	\subfloat[]{\includegraphics[width=.70\textwidth]{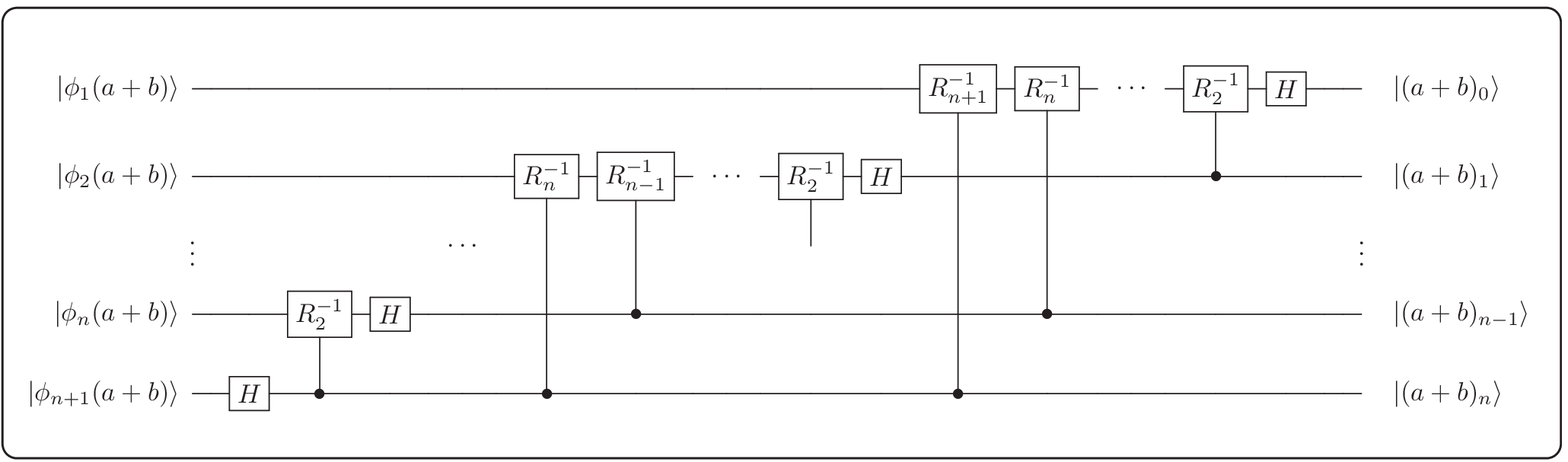}}%
	\hspace{1mm}%
	\subfloat[]{\includegraphics[width=.27\textwidth]{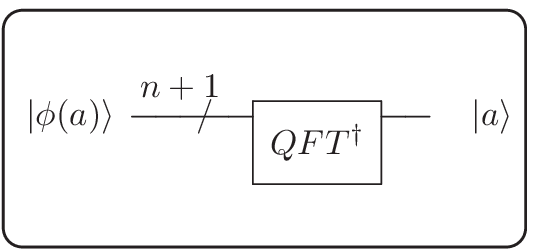}}}%
	\vspace*{8pt}%
	\caption{The inverse quantum Fourier transform (a) the quantum circuit of $QFT^{\dagger}$, (b) the simplified graph of $QFT^{\dagger}$.}%
	\label{Fig:2}%
\end{figure}

In this study, an overflow qubit $\ket{0}$ is used in addition to $n$-qubit state for non-modular arithmetic operations. This overflow qubit is actually the MSB of the result, ie the sign qubit. The circuits of the QFT and the inverse QFT are given to use for operations in the following sections. Thus, the quantum circuit of the $QFT$ application to the $n$-qubit state $\ket{a}$ with an additional overflow qubit ($n+1$-qubit in total) is given in Fig.~\ref{Fig:1}. The quantum circuit of applying $QFT^{\dagger}$ to state $\ket{\phi(a+b)}$ is given in Fig.~\ref{Fig:2}. 

The number of operations required for the QFT of any $n$-qubit input is $n(n+1)/2$ and the time complexity is $O({n^2})$. Barenco et al. proposed approximate quantum Fourier transform (AQFT) in Ref.~\citen{Barenco1996} in the presence of decoherence to reduce the number of operations required for QFT from $n(n+1)/2$ to $(2n-\log_{2}{n})(\log_{2}{n}-1)/2\approx n\log_2{n}$. However, the AQFT is used in algorithms that include periodicity estimates. Therefore, the QFT is used instead of the AQFT in this study.%

\section{Quantum Circuits for Arithmetic Operations}
\label{sec:3}

In this section, both modular and non-modular addition, subtraction, multiplication, division and exponentiation arithmetic operations based on QFT are presented. In addition, two's complement, absolute value calculation and comparison operations are presented. All the arithmetic operations are performed on two signed integers with the different number of qubits ($n \times m$).%

Let's consider the quantum states $\ket{a}$ and $\ket{b}$ of the signed integers $n$-bit $a$ and $m$-bit $b$ to be used in the arithmetic operations presented in this section as follows.

\begin{eqnarray} 
& \ket{a} = \ket{a_1 a_2 \cdots a_{n}} = \ket{a_1} \otimes \ket{a_2} \otimes \cdots \otimes \ket{a_{n}}. \label{Eq:6}\\
& \ket{b} = \ket{b_1 b_2 \cdots b_{m}} = \ket{b_1} \otimes \ket{b_2} \otimes \cdots \otimes \ket{b_{m}}. \label{Eq:7}
\end{eqnarray} 

where $a_{i}, b_j \in \{0, 1\}$, $i=1,2,...,n$, $j=1,2,...,m$ and $\otimes$ is a tensor product. $a_1$ and $b_1$ are sign qubits of the signed integers.

The required number of operations specified for all the proposed quantum arithmetic operators is in terms of the number of basic quantum gates.

\subsection{QFT based addition operation}
\label{sec:3.1}

Ruiz-Perez and Garcia-Escartin proposed modular and non-modular QFT based adder in Ref.~\citen{Perez2017} by adding an additional qubit to the Draper's modular QFT adder in Ref.~\citen{Draper2000}. The modified circuit in Ref.~\citen{Perez2017} could perform non-modular and modular addition operations on unsigned (i.e. positive) integers ($n \times n$). These circuits could also perform modular addition for signed integers up to $d/2$ ($d$ is the number of modulo). In this section, the improved version of the QFT adder in Ref.~\citen{Perez2017} is presented for both modular and non-modular adder on all signed integers ($n \times m$) without a limit such as $d/2$. To accomplish this, an additional qubit $\ket{0}$ which will be the sign qubit of the result, is first added to the state $\ket{a}$ like in Ruiz-Perez and Garcia-Escartin's design. For the modular addition, no additional qubit is required for the proposed method in this study. All the proposed circuits in this study are shown for $n \geq m$.%

\begin{equation}
\ket{0} \otimes \ket{a}  = \ket{a_0a_1 \cdots a_{n}}=\ket{a}. \label{Eq:8}%
\end{equation}

In non-modular addition, when both numbers are positive or negative, the sign qubits  $\ket{a_1}$ and $\ket{b_1}$ affect the $\ket{a_0}$ smoothly at the end of process. However, if one of the numbers is negative and one is positive, the overflow qubit also affects the $\ket{a_0}$. The sign qubit $\ket{a_0}$ is incorrect at the end of the operation. Therefore, Ruiz-Perez and Garcia-Escartin limits the addition to $d/2$ with signed integers. To correct this and to make the addition without any limit, two Toffoli gates with control qubits $\ket{a_1}$ of $\ket{a}$ and $\ket{b_1}$ of $\ket{b}$ (one of the gates is $0$-controlled on $\ket{a_1}$, the other is $0$-controlled on $\ket{b_1}$) are applied to the qubit $\ket{a_0}$. Then the QFT circuit shown in Fig.~\ref{Fig:1} is applied to the state $\ket{a}$. The states $\ket{\phi_i(a)}$ represent the $i$th qubit of the phase state $\ket{\phi(a)}$.

\begin{figure}%
	\centerline{\subfloat[]{\includegraphics[width=.70\textwidth]{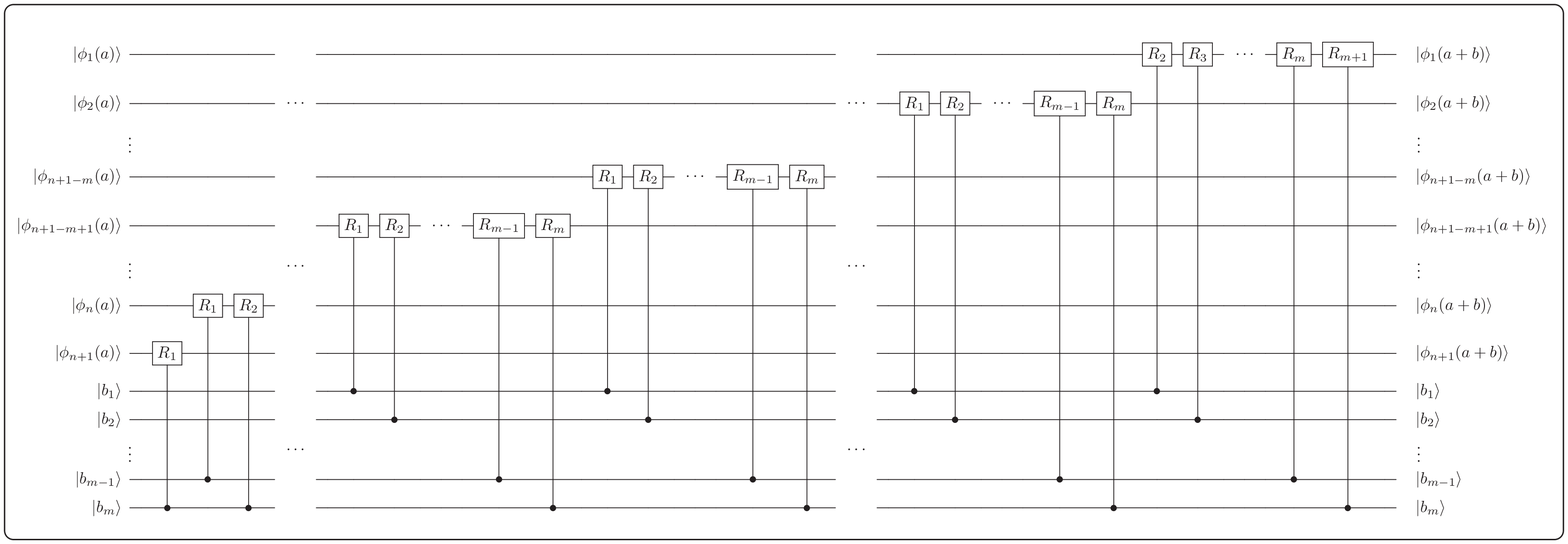}}%
	\hspace{1mm}%
	\subfloat[]{\includegraphics[width=.27\textwidth]{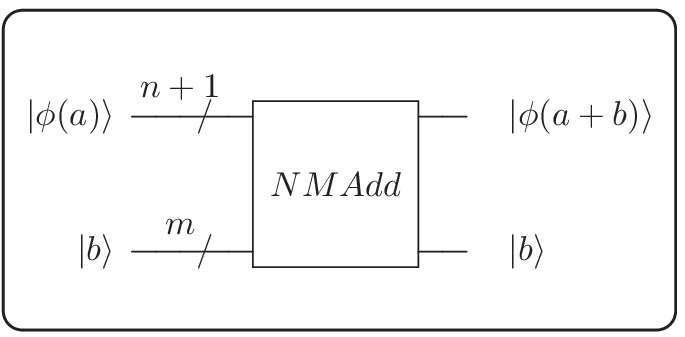}}}%
	\vspace*{8pt}%
	\caption{The non-modular addition with controlled rotation phase gates (NMAdd) (a) the quantum circuit of NMAdd, (b) the simplified graph of NMAdd.}%
	\label{Fig:3}%
\end{figure}

\begin{figure}%
	\centerline{
	\subfloat[]{\includegraphics[width=.70\textwidth]{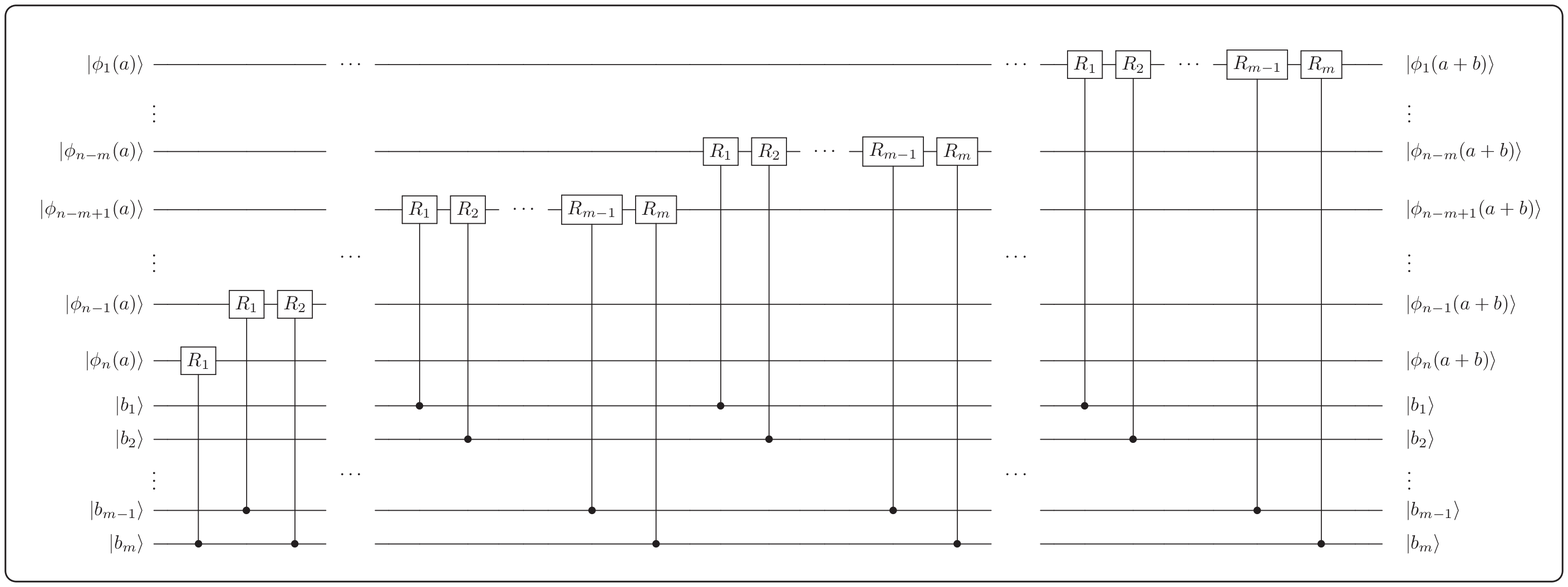}}%
	\hspace{1mm}%
	\subfloat[]{\includegraphics[width=.27\textwidth]{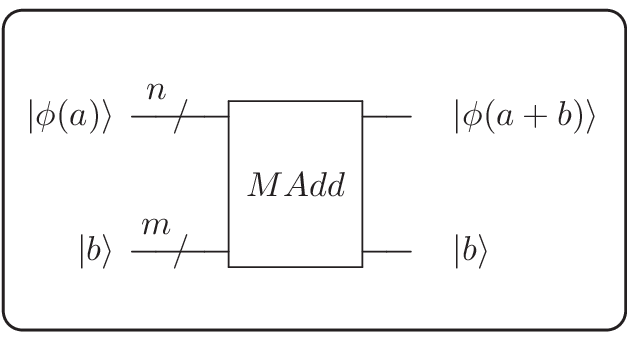}}}%
	\vspace*{8pt}%
	\caption{The modular addition with controlled rotation phase gates (MAdd) (a) the quantum circuit of MAdd, (b) the simplified graph of MAdd.}%
	\label{Fig:4}%
\end{figure}

In the Ruiz-Perez and Garcia-Escartin's QFT adder, the application of controlled rotation phase gates leads to error as a result of the adder process. The corrected circuit of adding state $\ket{b}$ to state $\ket{a}$ with controlled rotation phase gates for non-modular addition (NMAdd) is shown in Fig.~\ref{Fig:3}. Furthermore, the quantum circuit for modular addition (MAdd) is shown in Fig.~\ref{Fig:4}. The rotation phase gate is follows.

\begin{equation}
R_k=	\begin{bmatrix} 1 & 0 \\ 0 & e^{\frac{2\pi i}{2^k}} & \end{bmatrix}. \label{Eq:9}
\end{equation}

\begin{figure}%
	\centerline{
	\subfloat[]{\includegraphics[width=.50\textwidth]{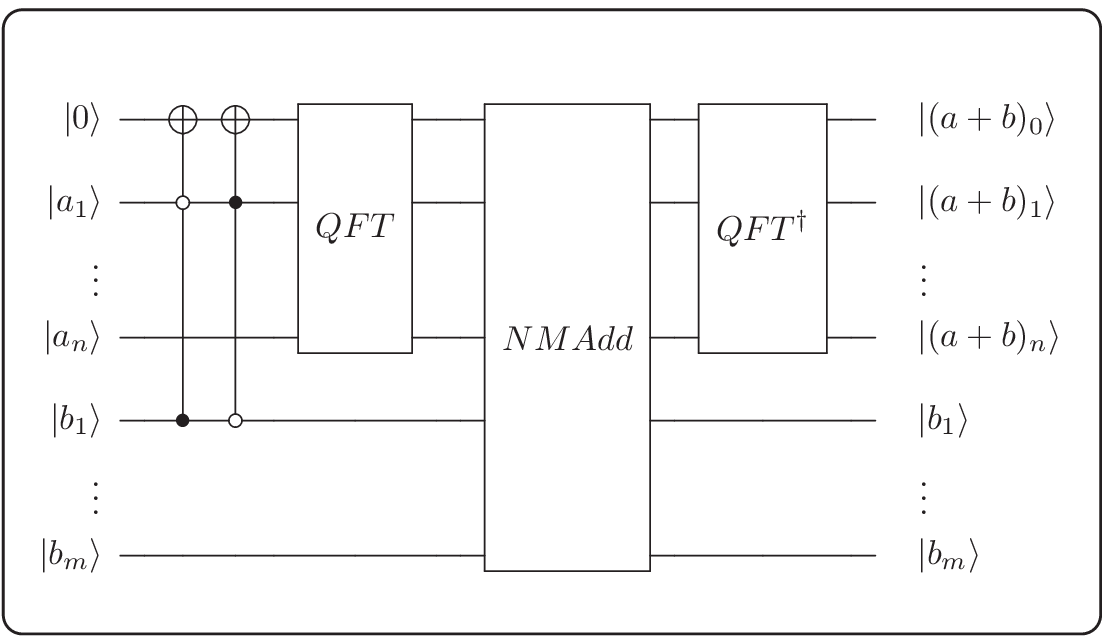}}%
	\hspace{1mm}%
	\subfloat[]{\includegraphics[width=.27\textwidth]{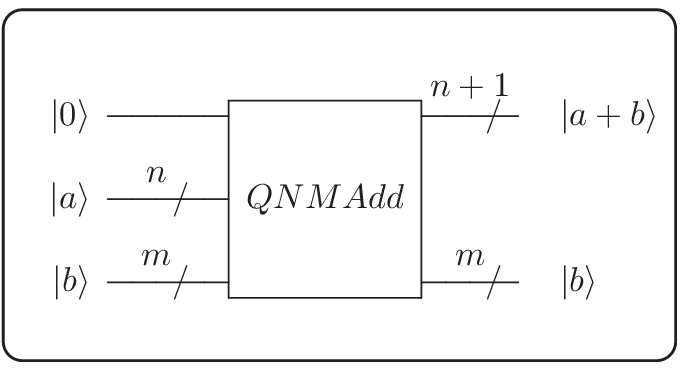}}}%
	\vspace*{8pt}%
	\caption{The non-modular QFT addition (QNMAdd) (a) the quantum circuit of QNMAdd, (b) the simplified graph of QNMAdd.}%
	\label{Fig:5}%
\end{figure}

\begin{figure}%
	\centerline{
	\subfloat[]{\includegraphics[width=.50\textwidth]{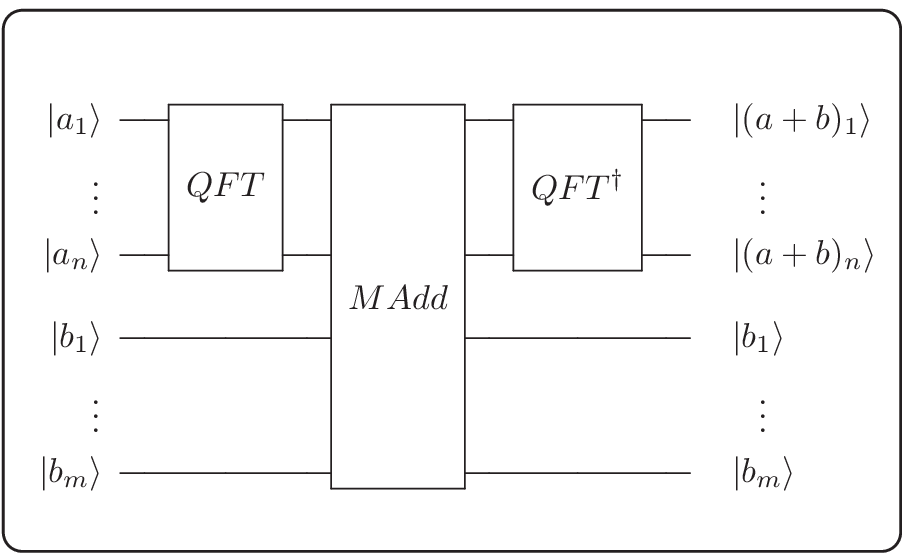}}%
	\hspace{1mm}%
	\subfloat[]{\includegraphics[width=.27\textwidth]{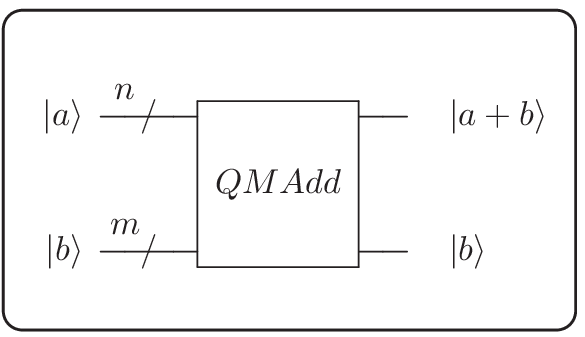}}}%
	\vspace*{8pt}%
	\caption{The modular QFT addition (QMAdd) (a) the quantum circuit of QMAdd, (b) the simplified graph of QMAdd.}%
	\label{Fig:6}%
\end{figure}

Finally, the state $\ket{a+b}$ is obtained by applying inverse QFT circuit shown in Fig.~\ref{Fig:2}. The full quantum circuits of QFT addition operation on the signed integers $n \times m$ are shown in Fig.~\ref{Fig:5} for non-modular addition (QNMAdd) and Fig.~\ref{Fig:6} for modular addition (QMAdd).

To add $m$-qubits signed integer $\ket{b}$ to $n$-qubits signed integer $\ket{a}$, $n+m$-qubits are used for modular addition, $n+m+1$-qubits are used for non-modular addition with only one additional qubit in this study. The proposed QFT adder requires no additional ancillary qubits compared to the other classic computation-based quantum adders and uses a minimum size of qubits for non-modular addition with signed integers ($n \times m$). A total of 16 basic gate operations (4 NOT and 2 Toffoli gates) are needed to correct the sign qubit before the QFT circuit. The number of operations required for the QFT circuit is $\left((n+1)(n+2)/2\right)$ for $(n+1)$-qubit input and the time complexity is $O(n^2)$. The number of operations required for the NMAdd circuit is $\left((m)(m+1)/2 + m(n+1-m)\right)=\left((n^2+n+1)/2\right)$ (for $n=m$) for the inputs $n$-qubits $\ket{a}$ and $m$-qubits $\ket{b}$ and the time complexity is $O(n^2)$. The number of operations required for the MAdd circuit is $\left((m)(m+1)/2 + m(n-m)\right)=\left((n^2+n)/2\right)$ (for $n=m$) for the inputs $n$-qubits $\ket{a}$ and $m$-qubits $\ket{b}$ and the time complexity is $O(n^2)$. The number of operations required for the inverse QFT circuit is $\left((n+1)(n+2)/2\right)$ for $(n+1)$-qubit input and the time complexity is $O(n^2)$. The number of operations required for the non-modular QFT addition (QNMAdd) circuit is $\left((n^2+3n+18) + (m)(2n-m+3)/2\right)=\left(3(n^2+3n+12)/2\right)$ (for $n=m$) for the inputs $n$-qubits $\ket{a}$ and $m$-qubits $\ket{b}$ and the time complexity is $O(n^2)$. The number of operations required for the modular QFT addition (QMAdd) circuit is $\left((n^2+n)+(m)(2n-m+1)/2\right)=\left(3(n^2+n)/2\right)$ (for $n=m$) for the inputs $n$-qubits $\ket{a}$ and $m$-qubits $\ket{b}$ and the time complexity is $O(n^2)$.%

\subsection{QFT subtraction operation}
\label{sec:3.2}

The QFT based subtraction operation is very similar to the QFT based addition operation. As with the QFT based addition operation, two NOT gates with zero and one controlled are initially applied for sign qubit correction, then the QFT is applied to the state $\ket{a}$. In the QFT subtraction operation, the inverse rotation phase gates are used instead of the rotation phase gates used in the QFT addition operation. The inverse rotation phase gate is follows.

\begin{equation}
R^{-1}_k=	\begin{bmatrix} 1 & & 0 \\ 0 & & e^{-\frac{2\pi i}{2^k}} & \end{bmatrix}. \label{Eq:10}%
\end{equation}

\begin{figure}%
	\centerline{
	\subfloat[]{\includegraphics[width=.70\textwidth]{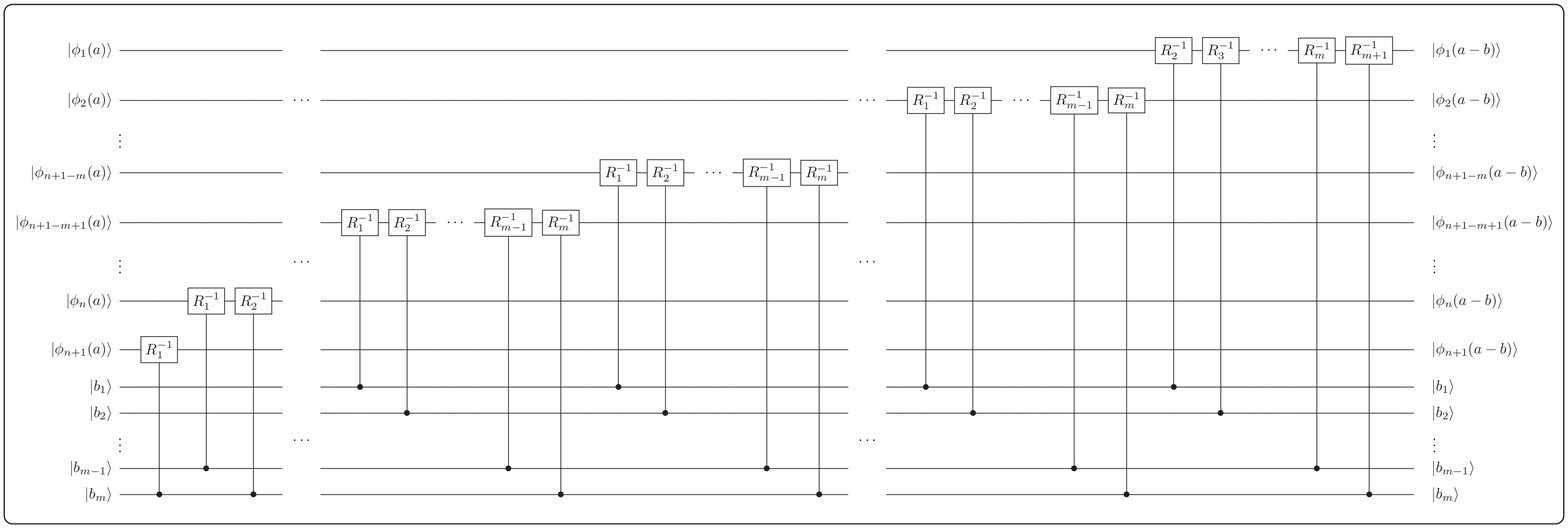}}%
	\hspace{1mm}%
	\subfloat[]{\includegraphics[width=.27\textwidth]{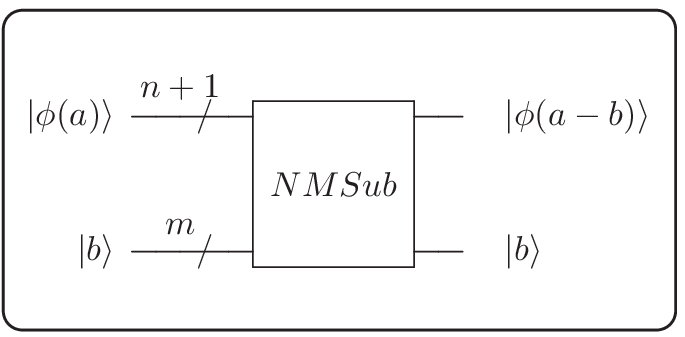}}}%
	\vspace*{8pt}%
	\caption{The non-modular subtraction with controlled rotation phase gates (NMSub) (a) the quantum circuit of NMSub, (b) the simplified graph of NMSub.}%
	\label{Fig:7}%
\end{figure}

\begin{figure}%
	\centerline{
	\subfloat[]{\includegraphics[width=.70\textwidth]{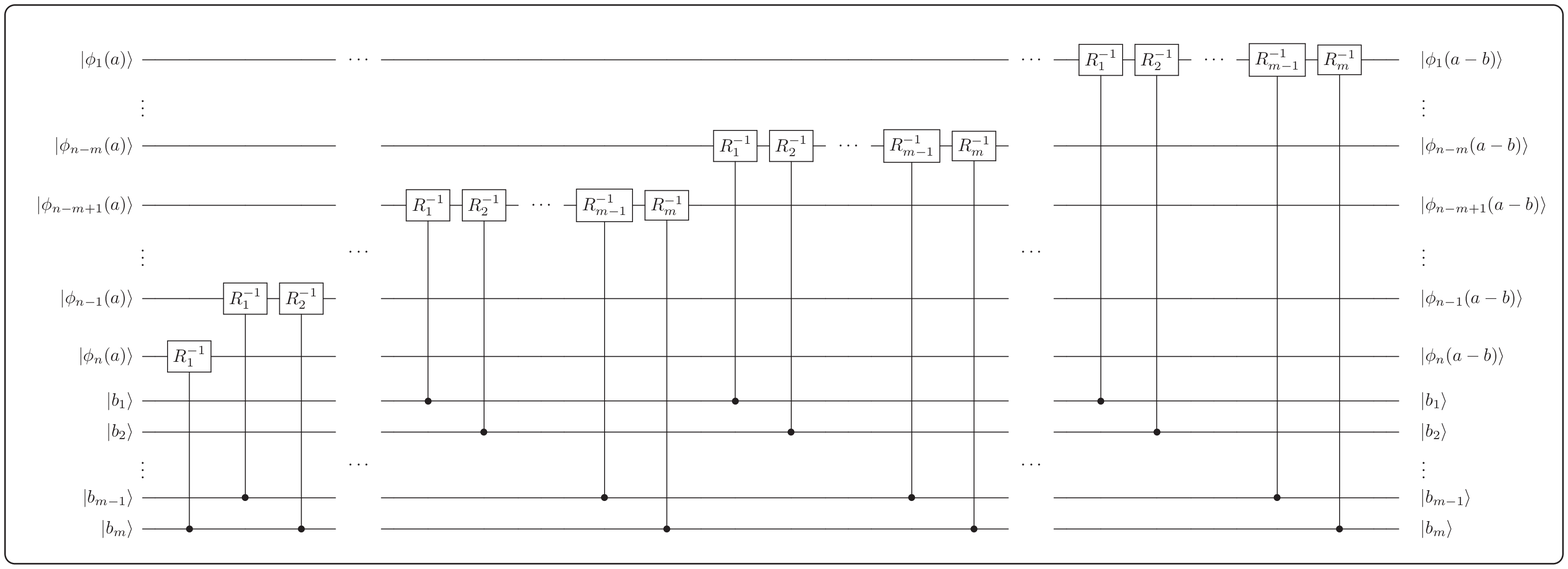}}%
	\hspace{1mm}%
	\subfloat[]{\includegraphics[width=.27\textwidth]{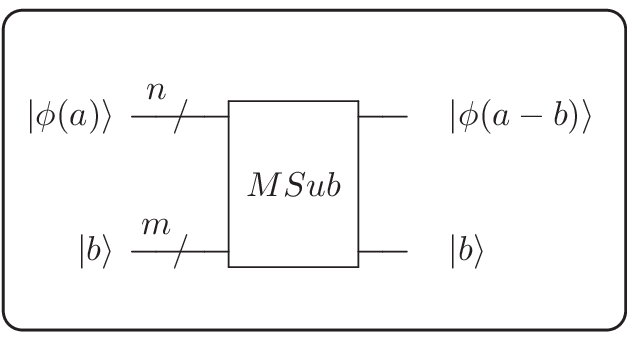}}}%
	\vspace*{8pt}%
	\caption{The modular subtraction with controlled rotation phase gates (MSub) (a) the quantum circuit of MSub, (b) the simplified graph of MSub.}%
	\label{Fig:8}%
\end{figure}

The circuit of subtracting state $\ket{b}$ from state $\ket{a}$ with controlled inverse rotation phase gates for non-modular subtraction (NMSub) is shown in Fig.~\ref{Fig:7}. The quantum circuit for modular subtraction (MSub) is shown in Fig.~\ref{Fig:8}.

\begin{figure}%
	\centerline{
	\subfloat[]{\includegraphics[width=.50\textwidth]{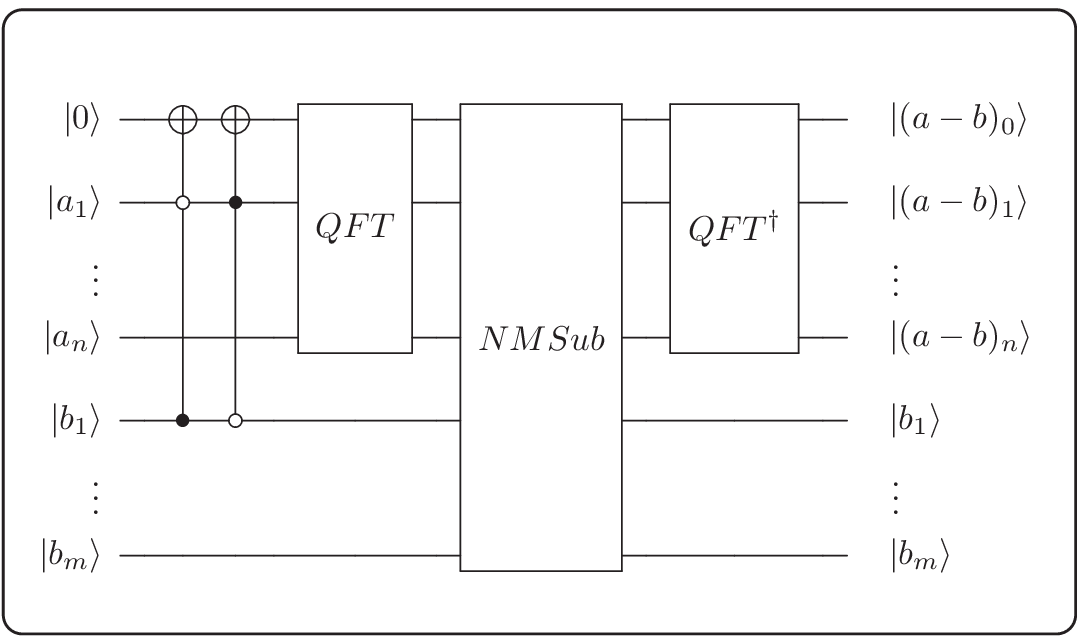}}%
	\hspace{1mm}%
	\subfloat[]{\includegraphics[width=.27\textwidth]{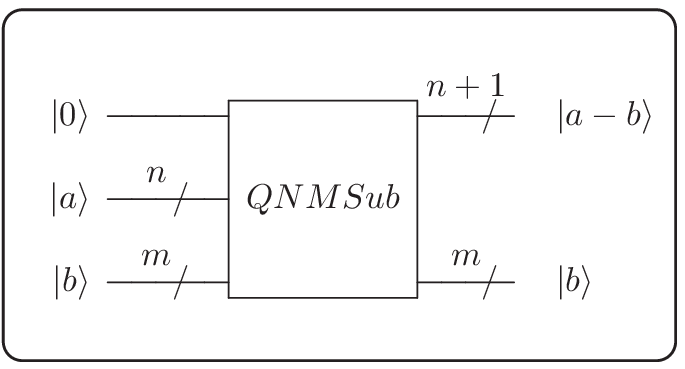}}}%
	\vspace*{8pt}%
	\caption{The non-modular QFT subtraction (QNMSub) (a) the quantum circuit of QNMSub, (b) the simplified graph of QNMSub.}%
	\label{Fig:9}%
\end{figure}

\begin{figure}%
	\centerline{
	\subfloat[]{\includegraphics[width=.50\textwidth]{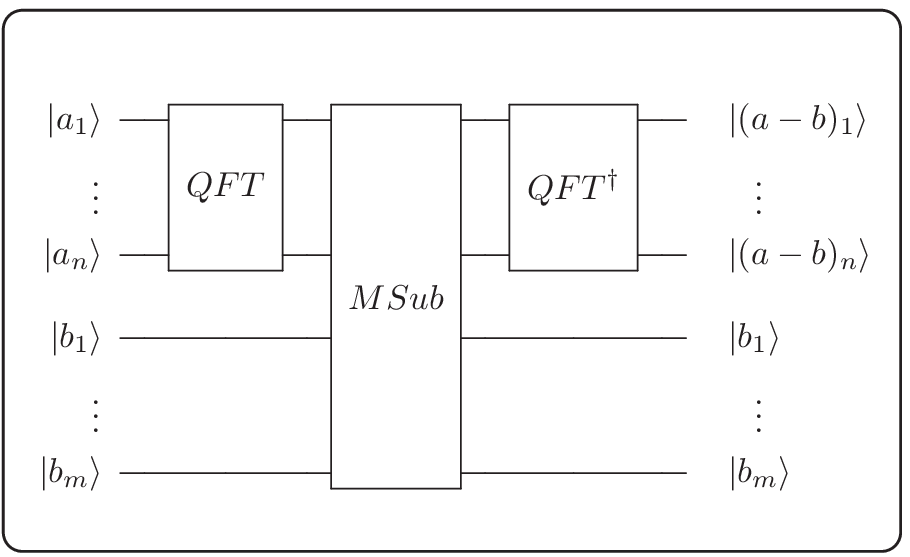}}%
	\hspace{1mm}%
	\subfloat[]{\includegraphics[width=.27\textwidth]{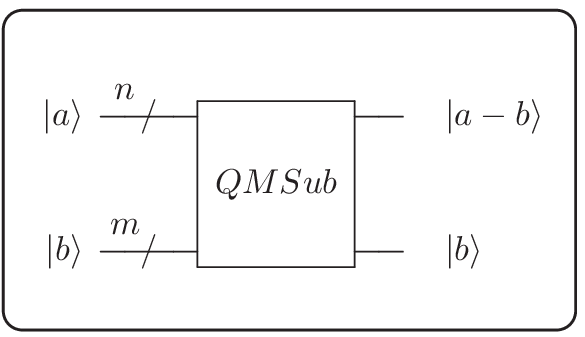}}}%
	\vspace*{8pt}%
	\caption{The modular QFT subtraction (QMSub) (a) the quantum circuit of QMSub, (b) the simplified graph of QMSub.}%
	\label{Fig:10}%
\end{figure}

Finally, the state $\ket{a-b}$ is obtained by applying inverse QFT circuit shown in Fig.~\ref{Fig:2}. The full quantum circuits of QFT subtraction operation on the signed integers $n \times m$ are shown in Fig.~\ref{Fig:9} for non-modular subtraction (QNMSub) and in Fig.~\ref{Fig:10} for modular addition (QMSub).

To subtract $m$-qubits signed integer $\ket{b}$ from $n$-qubits signed integer $\ket{a}$, $n+m$-qubits are used for modular addition, $n+m+1$-qubits are used with only one additional qubit for non-modular addition in this study. This QFT subtraction operation uses a minimum size of qubits for non-modular subtraction with signed integers ($n \times m$). The number of operations required for the NMSub circuit is $\left((m)(m+1)/2 + m(n+1-m)\right)=\left((n^2+3n)/2\right)$ (for $n=m$) for the inputs $n$-qubits $\ket{a}$ and $m$-qubits $\ket{b}$ and the time complexity is $O(n^2)$. The number of operations required for the MSub circuit is $\left((m)(m+1)/2 + m(n-m)\right)=\left((n^2+n)/2\right)$ (for $n=m$) for the inputs $n$-qubits $\ket{a}$ and $m$-qubits $\ket{b}$ and the time complexity is $O(n^2)$. The number of operations required for the non-modular QFT subtraction (QNMSub) circuit is $\left((n^2+3n+18)+(m)(2n-m+3)/2\right)=\left(3(n^2+3n+12)/2\right)$ (for $n=m$) for the inputs $n$-qubits $\ket{a}$ and $m$-qubits $\ket{b}$ and the time complexity is $O(n^2)$. The number of operations required for the modular QFT subtraction (QMSub) circuit is $\left((n^2+n)+(m)(2n-m+1)/2\right)=\left(3(n^2+n)/2\right)$ (for $n=m$) for the inputs $n$-qubits $\ket{a}$ and $m$-qubits $\ket{b}$ and the time complexity is $O(n^2)$.

To find the difference of one number from another number, the addition operation can be done with the opposite of the number instead of the subtraction operation. In other words, the operation $a+(-b)$ can also be taken into consideration to find the result of the operation $a-b$. To find the difference of the two numbers by the opposite of the number, firstly ($-b$) is found with the proposed quantum two’s complement circuit (QTC) shown in Fig.~\ref{Fig:11} in this study, then $a+(-b)$ is found with the proposed addition circuit. Thus, to find the result of subtraction by adding the opposite of the number, in order to find the opposite of the $n$-qubit number ($-b$) with the QTC circuit, an additional $1$ ancillary qubit and ($n^2+n$) operations are also required in addition to the QMAdd or QNMAdd modules (according to modular or non-modular subtraction). Therefore, the number of operations to be applied and the time complexity increase with the opposite number operation. The time complexities and the number of the operations required for QFT addition ($a+b$) and QFT subtraction ($a-b$) operations proposed in this study are equal. The only difference between them is the use of $R_k$ gates in the addition operation and the $R_k^{-1}$ gates in the subtraction operation. 

\subsection{QFT based absolute value operation}
\label{sec:3.3}
\begin{figure}%
	\centerline{
	\subfloat[]{\includegraphics[width=.42\textwidth]{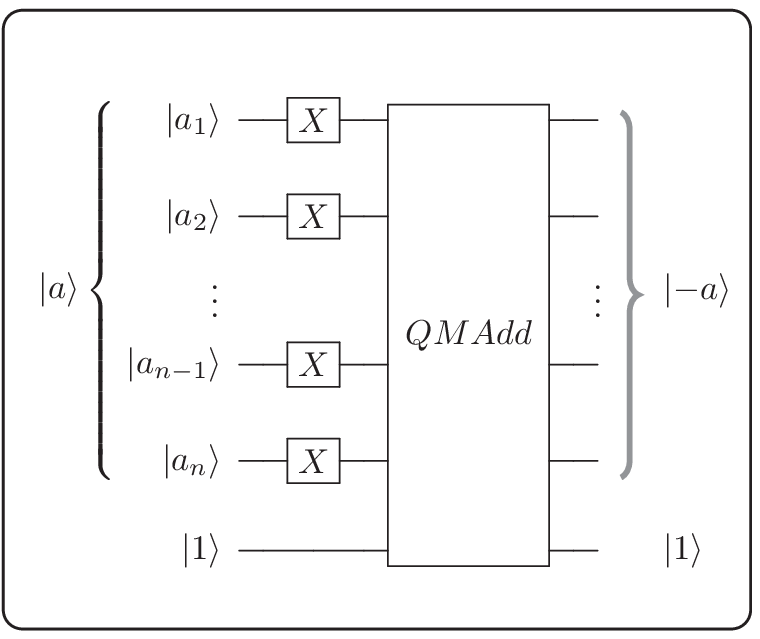}}%
	\hspace{1mm}%
	\subfloat[]{\includegraphics[width=.25\textwidth]{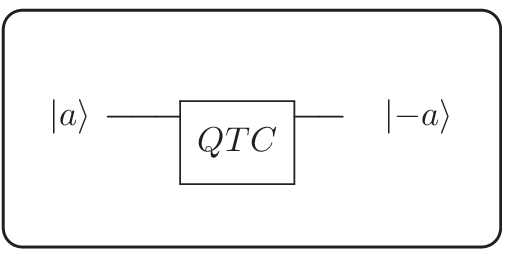}}}%
	\vspace*{8pt}%
	\caption{The two's complement of signed binary integer (QTC) (a) the quantum circuit of QTC, (b) the simplified graph of QTC.}%
	\label{Fig:11}%
\end{figure}

Absolute value is a positive expression of a number. In other words, the negative number is multiplied by -1. The two's complement method is used to change the sign by multiplying a number with -1 in the binary number system. If the number is negative, the number is converted to positive by the two's complement method and thus its absolute value is found. The quantum circuit of the two's complement operation is required to inverse the sign of a quantum state representing a signed integer. This quantum complement circuit can be used in all kinds of sign changing operations, primarily absolute value calculation. The quantum circuit of the two's complement method (QTC) for $n$-qubits $\ket{a}$ ($\ket{a_1}$ is a sign qubit) is shown in Fig.~\ref{Fig:11}.

\begin{figure}%
	\centerline{
	\subfloat[]{\includegraphics[width=.4\textwidth]{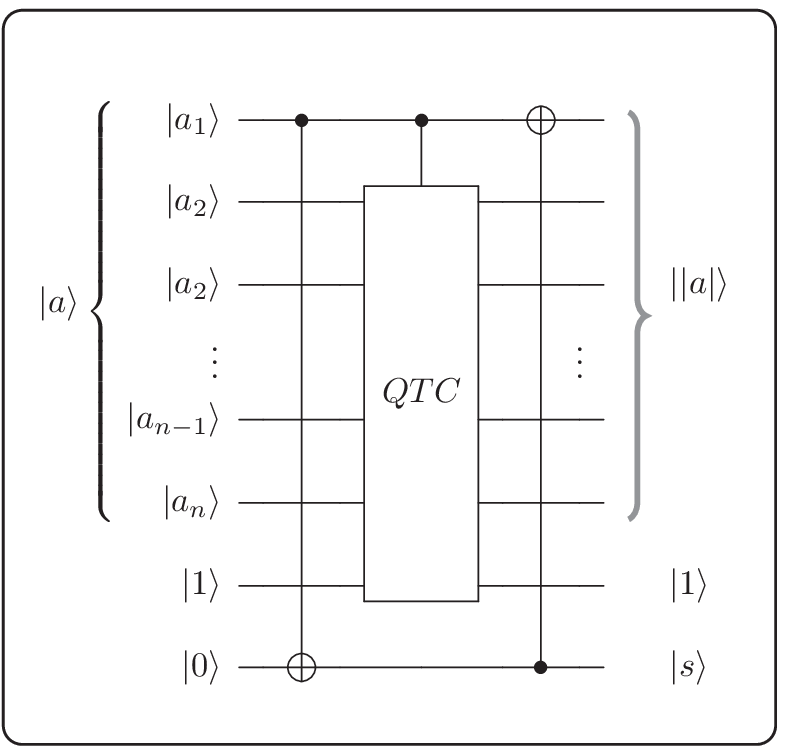}}%
	\hspace{1mm}%
	\subfloat[]{\includegraphics[width=.27\textwidth]{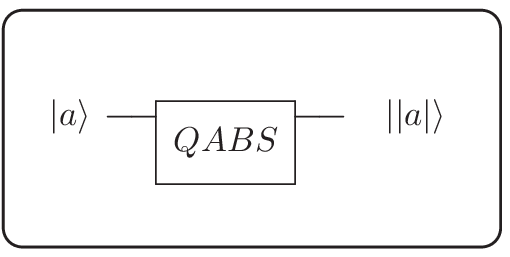}}}%
	\vspace*{8pt}%
	\caption{The QFT based absolute value (QABS) (a) the quantum circuit of QABS, (b) the simplified graph of QABS.}%
	\label{Fig:12}%
\end{figure}

Where $\ket{1}$ is for adding 1 in the two's complement method. If the sign qubit $\ket{a_1}=0$, $\ket{a}$ is a positive number and no action is required for the absolute value. If the sign qubit $\ket{a_1}=1$, $\ket{a}$ is negative and the two's complement must be applied to $\ket{a}$. First, an additional ancillary qubit $\ket{0}$ is required to store the sign qubit $\ket{a_1}$ of the state $\ket{a}$. Then by applying CNOT($\ket{a_1}$,$\ket{0}$) gate, the state $\ket{a_1}$ is transferred to the state $\ket{0}$ as $\ket{s}$. The controlled ($\ket{a_1}$) QTC circuit is applied to other qubits of the state $\ket{a}$ ($\ket{a_2...a_n}$) and state $\ket{1}$. Finally, by applying CNOT($\ket{s}$,$\ket{a_1}$) gate, if the number is negative, the sign qubit $\ket{a_1}$ changes from $\ket{1}$ to $\ket{0}$, so the number will be positive. The quantum circuit of the absolute value operation is shown in Fig.~\ref{Fig:12}.

The QTC operation requires a total of $n+1$-qubits with an ancillary qubit for an input $n$-qubits. The QABS operation requires a total of $n+2$-qubits with two ancillary qubits for an input $n$-qubits. The number of operations required for the QTC circuit is $\left(n^2+3n\right)$ for the input $n$-qubits $\ket{a}$ and the time complexity is $O(n^2)$. The number of operations required for the QABS circuit is $\left(n^2+3n+2\right)$ for the input $n$-qubits $\ket{a}$ and the time complexity is $O(n^2)$.

\subsection{QFT based comparison operation}
\label{sec:3.4}

The quantum comparator (QC) module proposed by Wang et al. \cite{Wang2012} is used for comparison of quantum states in most of the studies. The QC module in Ref.~\citen{Wang2012} compares two $n$-qubits unsigned integers (positive) with the same number of qubits. In this study, the proposed QFT based quantum comparison (QComp) scheme can be used to compare all signed integers with different numbers of qubits ($n \times m$). The QC module in Ref.~\citen{Wang2012} uses a lot of qubits and has a lot of time complexity. 

\begin{figure}%
	\centerline{
	\subfloat[]{\includegraphics[width=.5\textwidth]{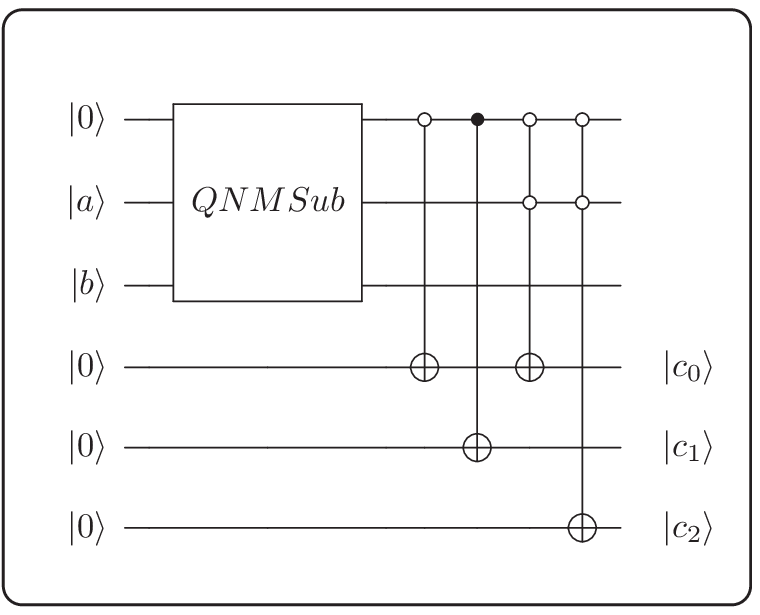}}%
	\hspace{1mm}%
	\subfloat[]{\includegraphics[width=.37\textwidth]{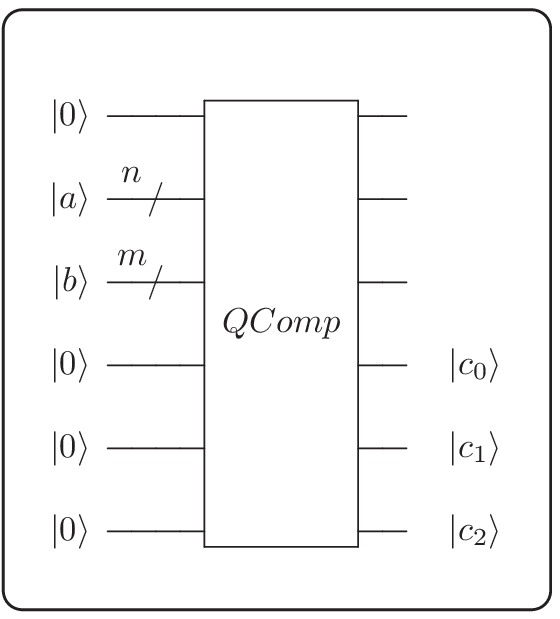}}}%
	\vspace*{8pt}%
	\caption{The QFT based comparison operation (QComp) (a) the quantum circuit of QComp, (b) the simplified graph of QComp.}%
	\label{Fig:13}%
\end{figure}

The proposed QComp scheme can be compare all signed integers using the QFT-based QNMSub operation and 3 ancillary qubits $\ket{0}$. The proposed method is quite simple and based on the subtraction of the signed integer state $\ket{b}$ from the signed integer state $\ket{a}$. The quantum circuit of QComp is shown in Fig.~\ref{Fig:13}. After subtraction from $\ket{a}$ to $\ket{b}$, one of the three ancillary quits changes to $\ket{1}$ by controlled NOT gates. According to the measurement results of qubits $c_0, c_1, c_2$; if $c_0=1$ then $a>b$, if $c_1=1$ then $a<b$, if $c_2=1$ then $a=b$.

The QC module in Ref.~\citen{Wang2012} needs additional $2n$ ancillary qubits to compare $n$-qubits state $\ket{a}$ with $n$-qubits state $\ket{b}$. The proposed QComp module only needs 3 additional ancillary qubits for the result. The QC module uses two times (01)-controlled NOT and (10)-controlled NOT gates, $\frac{1}{2}n(n-1)$ times (0010)-controlled NOT and (0001)-controlled NOT gates for comparison. Since the n-CNOT gate consists of $2n-1$ Toffoli gates and 1 CNOT gate, the scheme in Ref.~\citen{Wang2012} uses a lot of basic gates. The number of operations required for the QC in Ref.~\citen{Wang2012} is $\left(48n^2-48n+16\right)$ for $n \times n$ inputs. The number of operations required for the proposed QComp is $\left(n^2+3n+41+m(2n-m+3)/2\right)$ for $n \times m$ inputs $(\text{for } n=m,$  $\left((3n^2+9n+82)/2\right))$ and the time complexity is $O(n^2)$.%

\subsection{QFT based multiplication operation}
\label{sec:3.5}

The QFT multiplier proposed in Ref.~\citen{Perez2017} performs non-modular multiplication of unsigned integers ($n \times n$).  Alvarez-Sanchez et al. proposed quantum multiplication on signed integers ($n \times n$) in Ref.~\citen{Sanchez2008} based on the Booth algorithm which is widely used in classical computer. The result is stored in $2n$-qubits separately from the input qubits in Alvarez-Sanchez et al.'s circuit. In this section, the quantum circuit of a novel non-modular QFT-based multiplication is proposed which multiplies the signed integers of all different qubit numbers ($n \times m$). The proposed circuit also needs $n+m-1$-qubits (for $n=m$, $2n-1$-qubits) for the multiplication result $\ket{a.b}$ of $n$-qubits $\ket{a}$ and $m$-qubits $\ket{b}$.

\begin{figure}%
	\centerline{
	\subfloat[]{\includegraphics[width=.45\textwidth]{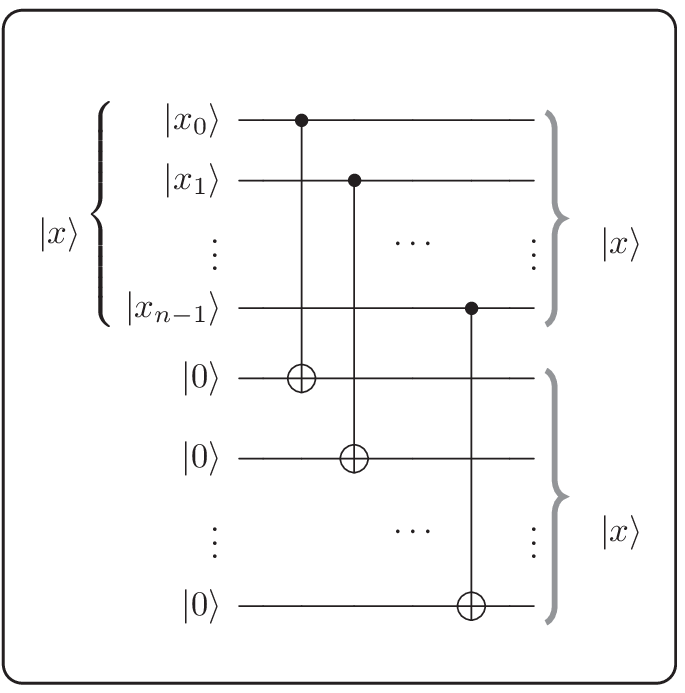}}%
	\hspace{1mm}%
	\subfloat[]{\includegraphics[width=.27\textwidth]{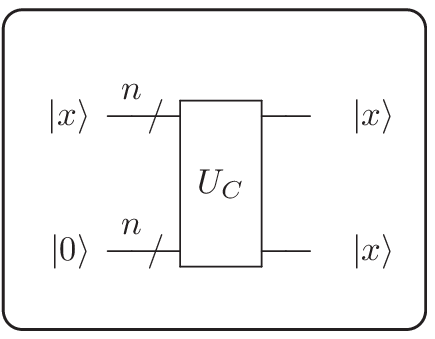}}}%
	\vspace*{8pt}%
	\caption{The Unitary Copy ($U_C$) (a) the quantum circuit of $U_C$, (b) the simplified graph of $U_C$.}%
	\label{Fig:14}%
\end{figure}

In order not to change the values of inputs $\ket{a}$ and $\ket{b}$ in the process, copies of the inputs are made in the proposed circuit and operations are performed on them. This is done by taking into consideration the studies that need to reuse the input states. The copy states are treated as inputs and multiplication occurs in these copy states. If it is not important to change the values of inputs, the operations in the circuit can be directly processed on states $\ket{a}$ and $\ket{b}$. In this case, no extra $m$ ancillary qubits will be required. The unitary $U_C$ operation for copying an $n$-qubits state $\ket{x}$ to another $n$-qubits state $\ket{0}$ is shown in Fig.~\ref{Fig:14}.

\begin{equation}
U_C\ket{x}\ket{0} = \ket{x}\ket{x}. \label{Eq:11}%
\end{equation}

\begin{figure}%
	\centerline{
	\subfloat[]{\includegraphics[width=.75\textwidth]{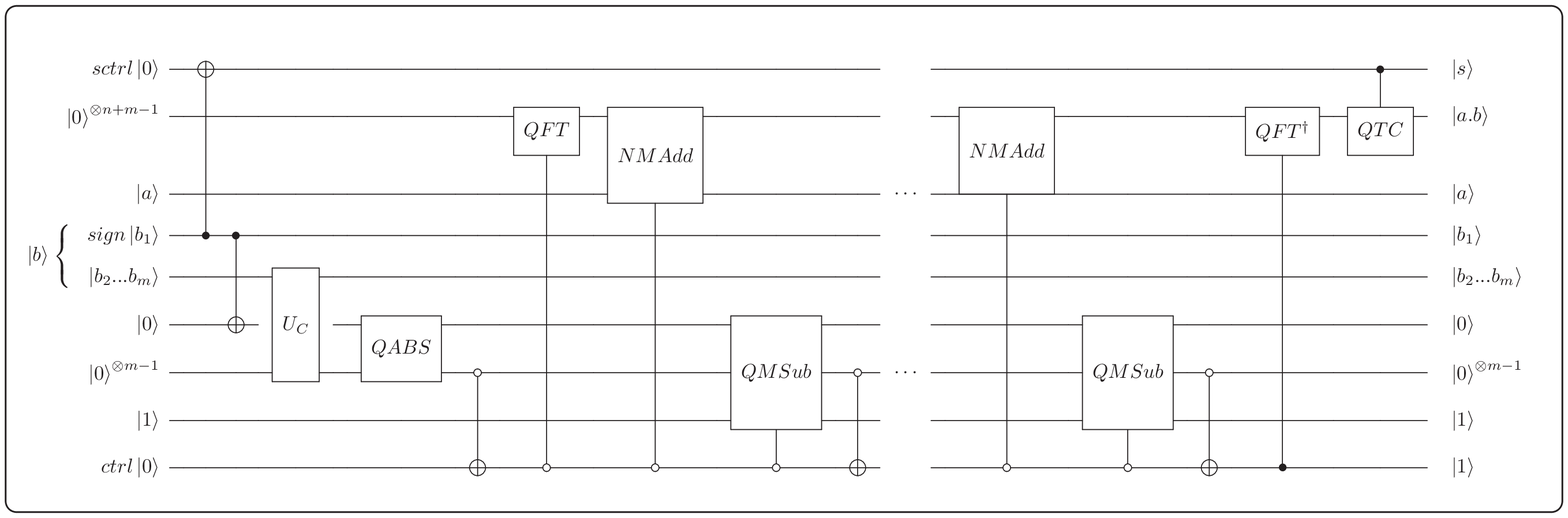}}%
	\hspace{1mm}%
	\subfloat[]{\includegraphics[width=.22\textwidth]{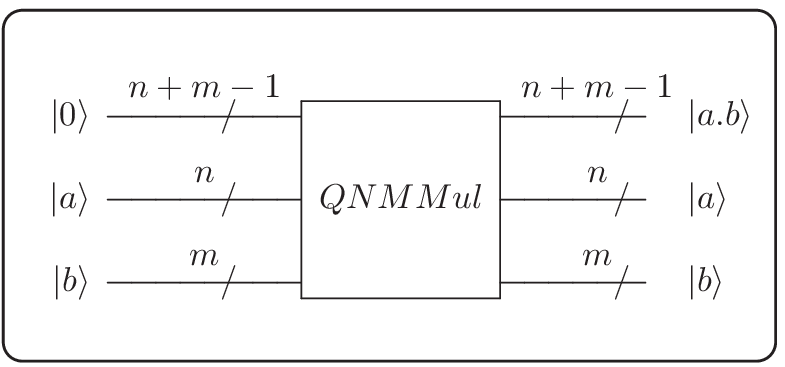}}}%
	\vspace*{8pt}%
	\caption{The QFT based multiplication operation (QNMMul) (a) the quantum circuit of QNMMul, (b) the simplified graph of QNMMul. The flaw of the operation $U_C$ indicates that $U_C$ is not applied to the qubit $\ket{0}$.}%
	\label{Fig:15}%
\end{figure}

The result of the $a.b$ operation is found by $|b|$ (absolute value of $b$) times adding $a$ value to $0$ in this study. If $\ket{b}$ is a negative integer, its absolute value will be positive. Therefore, the signal qubit of the result will be incorrect. Since the operation is performed with the absolute value of $\ket{b}$, the qubit $\ket{0}$ called $sctrl$ is defined to correct the sign of the result $\ket{a.b}$. The sign qubit $\ket{b_1}$ of state $\ket{b}$ is transferred to the $sctrl$ state $\ket{0}$ with CNOT gate. If $\ket{b}$ is a negative integer, $\ket{b_1}$ is $\ket{1}$ and $sctrl$ will be $\ket{1}$. Hence, at the end of the procedure, the result is corrected by applying $scrtl$-controlled the two's complement (QTC) circuit. After the signal information is transferred to the $sctrl$ qubit, the state $\ket{b}$ is backed up to the ancillary qubits $\ket{0}^{\otimes m}$ as $\ket{b^{'}}$ with the operation $U_C$. The absolute value of $\ket{b}$ is calculated on the ancillary qubits $\ket{b^{'}}$. The QFT is applied to the allocated $n+m-1$-qubits for the result. Then $\ket{a}$ is added to this result qubits by NMAdd, $\ket{1}$ is subtracted from $\ket{b^{'}}$ by QMSub and a zero-controlled NOT gate is applied to the ancillary qubit $\ket{0}$ as $ctrl$ to check if state $\ket{b^{'}}=\ket{0}^{\otimes m-1}$. If the state $ctrl=\ket{1}$, the process is terminated by applying the inverse QFT and the controlled QTC. The quantum circuit of QFT-based multiplication (QNMMul) on two signed integers ($n \times m$) is shown in Fig.~\ref{Fig:15}.

The proposed QFT-based multiplication operation requires $m+n-1$-qubits for the result $\ket{a.b}$, $m$-qubits for the copy of the state $\ket{b}$ (If the change of $\ket{b}$ is not important, these qubits are not needed) and 3 ancillary qubits ($\ket{0}$ as sctrl, $\ket{1}$, $\ket{0}$ as ctrl). The time complexity of the $U_C$ circuit is $O(n)$ for $n$-qubits input. The number of operations required for the QNMMul circuit is $\left((5n^2+n)/2+4m^2+4nm+6m+7\right)=\left((21n^2+13n)/2+7\right)$ (for $n=m$) for the inputs $n$-qubits $\ket{a}$ and $m$-qubits $\ket{b}$ and the time complexity is $O(n^2)$.

\subsection{QFT based division operation}
\label{sec:3.6}
In this section, the quantum circuit of a new non-modular QFT-based division is proposed which divides the signed integers of all different qubit numbers ($n \times m$). The proposed circuit also needs $n$-qubits for the division result $\ket{a/b}$ of $n$-qubits $\ket{a}$ and $m$-qubits $\ket{b}$.

In the division operation, as in the multiplication operation, the copies of the inputs $\ket{a}$, $\ket{b}$ are made in order not to change the inputs. The copy states are treated as inputs and division occurs in these copy states. If it is not important to change the values of inputs, the operations in the circuit can be directly processed on states $\ket{a}$ and $\ket{b}$. In this case, no extra $n+m$ ancillary qubits will be required. 

\begin{figure}%
	\centerline{
	\subfloat[]{\includegraphics[width=.75\textwidth]{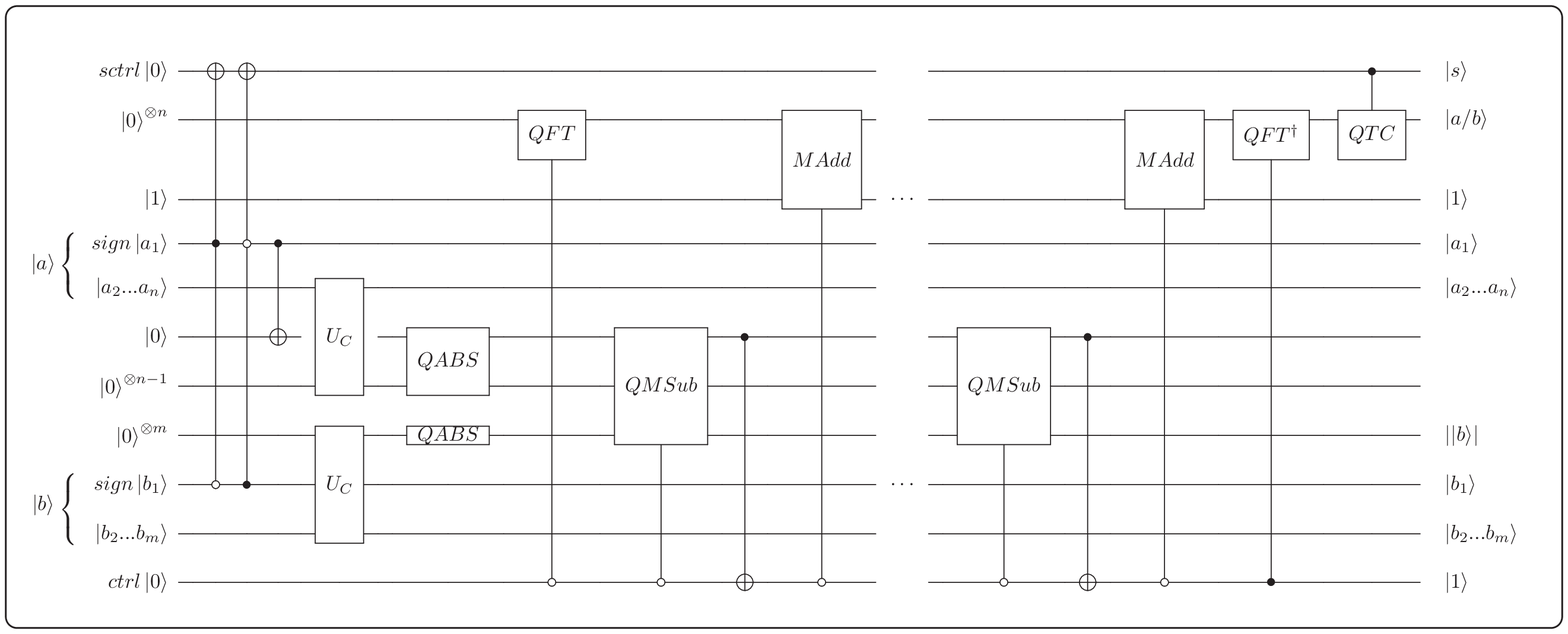}}%
	\hspace{1mm}%
	\subfloat[]{\includegraphics[width=.22\textwidth]{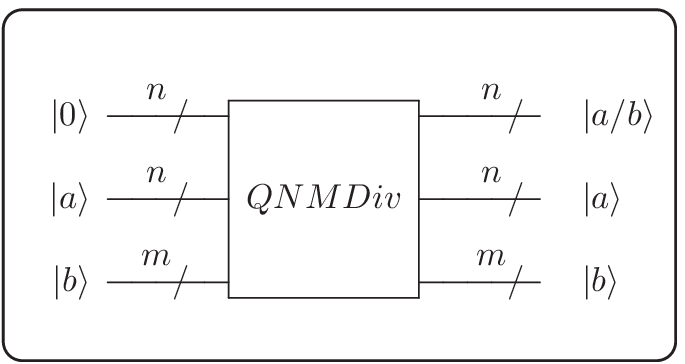}}}%
	\vspace*{8pt}%
	\caption{The QFT based division operation (QNMDiv) (a) the quantum circuit of QNMDiv, (b) the simplified graph of QNMDiv. The flaw of the operation $U_C$ indicates that $U_C$ is not applied to the qubit $\ket{0}$.}%
	\label{Fig:16}%
\end{figure}

The result of the $a/b$ operation is obtained by taking the absolute values $|a|, |b|$ of $a$ and $b$ first, and the number of steps of subtraction of $a$ from $b$ until the value of $a$ is negative. If only one of the $\ket{a}$ and the $\ket{b}$ is negative, its absolute values will be positive. Therefore, the signal qubit of the result will be incorrect. Since the operation is performed with the absolute value of $\ket{a}$ or $\ket{b}$, the qubit $\ket{0}$ called $sctrl$ is defined to correct the sign of the result $\ket{a/b}$. If the signs of the input states $\ket{a_1}$ and $\ket{b_1}$ are different, NOT gates are applied to the $sctrl$ qubit with 01-controlled and 10-controlled at the inputs and the sign information stored in the $strcl$ qubit for the correction of the result's sign at the end of the operations. If only one of the $\ket{a}$ and the $\ket{b}$ is a negative integer, $\ket{b_1}$ or $\ket{b_1}$ is $\ket{1}$ and $sctrl$ will be $\ket{1}$. Hence, at the end of the procedure, the result is corrected by applying $scrtl$-controlled the two's complement (QTC) circuit. After the signal information is transferred to the $sctrl$ qubit, the states $\ket{a}$ and $\ket{b}$ are backed up to the ancillary qubits $\ket{0}^{\otimes n}$ and $\ket{0}^{\otimes m}$ as $\ket{a^{'}}$ and $\ket{b^{'}}$ with the operation $U_C$. The absolute values of $\ket{a}$ and $\ket{b}$ are calculated on the ancillary qubits $\ket{a^{'}}$ and $\ket{b^{'}}$. The QFT is applied to the allocated $n$-qubits for the result. Then $\ket{b^{'}}$ is subtracted from the $\ket{a^{'}}$ by QMSub, CNOT gate (controlled qubit:$\ket{{a_0}^{'}}$) is applied to the $ctrl\ket{0}$ qubit to check if state $\ket{{a_0}^{'}}=\ket{1}$ and $\ket{1}$ is added to $\ket{a/b}$ by MAdd. If the state $ctrl=\ket{1}$, the process is terminated by applying the inverse QFT and the controlled QTC. The quantum circuit of QFT-based division (QNMDiv) on two signed integers ($n \times m$) is shown in Fig.~\ref{Fig:16}.

The proposed QFT-based division operation requires $n$-qubits for the result $\ket{a/b}$, $n$-qubits for the copy of the state $\ket{a}$ and $m$-qubits for the copy of the state $\ket{b}$ (If the changes of $\ket{a}$ and $\ket{b}$ are not important, these qubits are not needed) and 3 ancillary qubits ($\ket{0}$ as sctrl, $\ket{1}$, $\ket{0}$ as ctrl). The number of operations required for the QNMDiv circuit is $\left(3n^2+12n+nm+(m^2+11m)/2+36\right)=\left((9n^2+35n)/2+36\right)$ (for $n=m$) for the inputs $n$-qubits $\ket{a}$ and $m$-qubits $\ket{b}$ and the time complexity is $O(n^2)$.

\subsection{QFT based exponentiation operation}
\label{sec:3.7}

In this section, the quantum circuit of non-modular QFT based exponentiation operation for signed integers ($n \times m$) is proposed. All the operations presented in this study produce results as a signed integer by taking signed integers as inputs. The exponentiation operation presented in this section can be applied to all situations where the value of exponent$\geq 0$. Because we can express the operation $a^{-b}$ as $1/a^b$ for $-b<0$ and the result is a decimal number between $0$ and $1$.

We can express the exponentiation operation $a^b$ as multiplying the base number $a$ by itself the exponent $b$ times, and it is shown as follows. 

\begin{equation}
a^b = \underbrace{a \times a \times \cdots \times a}_{b}
\label{Eq:12}
\end{equation}

\begin{figure}%
	\centerline{
		\subfloat[]{\includegraphics[width=.72\textwidth]{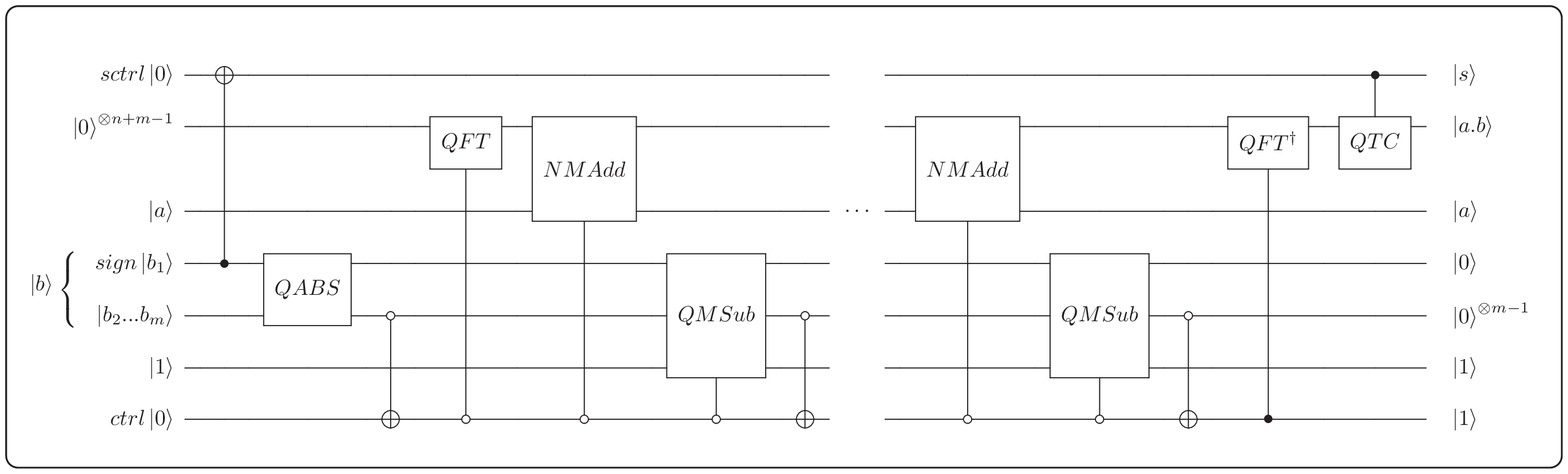}}%
		\hspace{1mm}%
		\subfloat[]{\includegraphics[width=.25\textwidth]{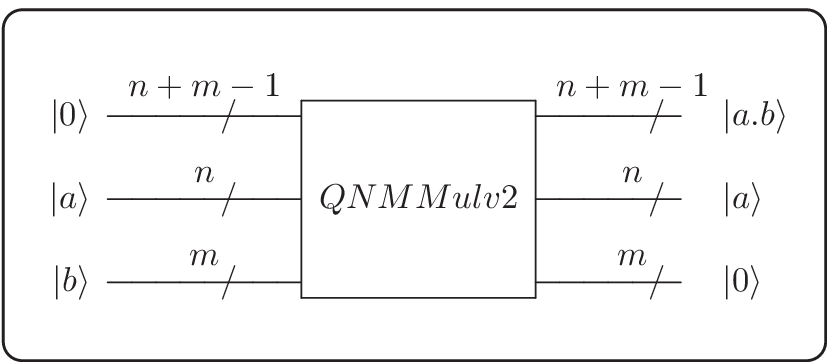}}}%
	\vspace*{8pt}%
	\caption{The QFT based multiplication operation (QNMMulv2) on the original input states (a) the quantum circuit of QNMMulv2, (b) the simplified graph of QNMMulv2.}%
	\label{Fig:17}%
\end{figure}

The quantum exponentiation operation can be performed using the arithmetic operators described in the above sections. The copies of the inputs are taken to the ancillary qubits and the operations are performed on the ancillary qubits in order to prevent the inputs from changing in the multiplication and division operations presented in the previous sections. If the change of input values is insignificant, it is also stated in the previous sections that the operations on copy ancillary qubits can be applied directly on the input qubits. Thus, no additional ancillary qubits will be needed. In this context, the QNMMulv2 which is modified version of the simplified graph of QNMMul shown in Fig.~\ref{Fig:15}, is shown in Fig.~\ref{Fig:17}. Here, the operations are applied directly on the input states $n$-qubits $\ket{a}$ and $m$-qubits $\ket{b}$.The state $\ket{a}$ does not change, but the state $\ket{b}$ changes and becomes $\ket{0}^{\otimes m}$ at the end of the multiplication operation. The result of the multiplication is stored in the $n+m-1$-ancillary qubits.

\begin{figure}%
	\centerline{
		\subfloat[]{\includegraphics[width=.32\textwidth]{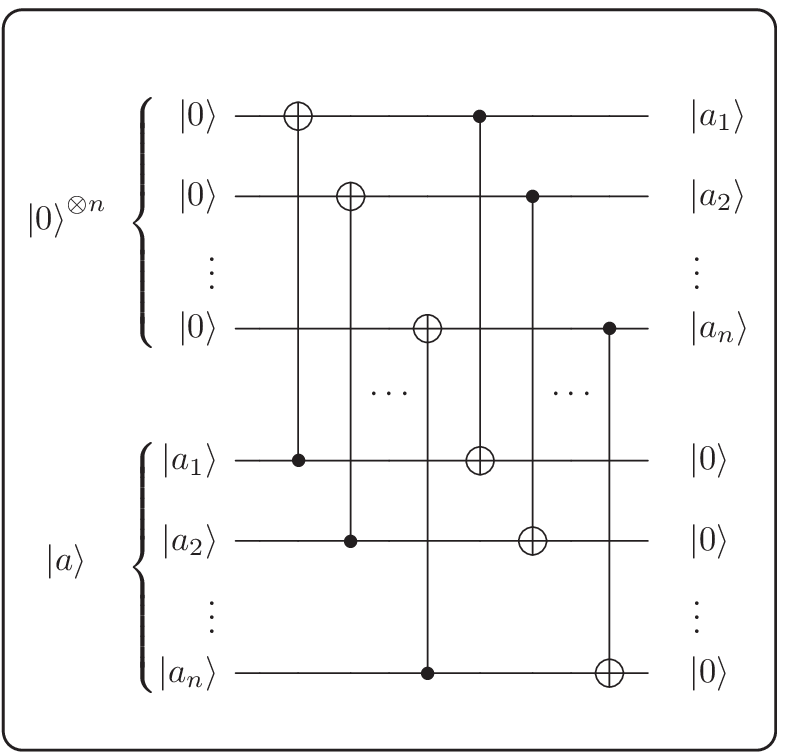}}%
		\hspace{1mm}%
		\subfloat[]{\includegraphics[width=.25\textwidth]{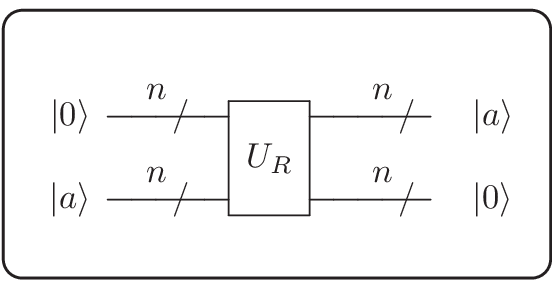}}}%
	\vspace*{8pt}%
	\caption{The operation $U_R$ that changes any state with the state $\ket{0}$ (a) the quantum circuit of $U_R$, (b) the simplified graph of $U_R$.}%
	\label{Fig:18}%
\end{figure}

\begin{figure}%
	\centerline{
		\subfloat[]{\includegraphics[width=.7\textwidth]{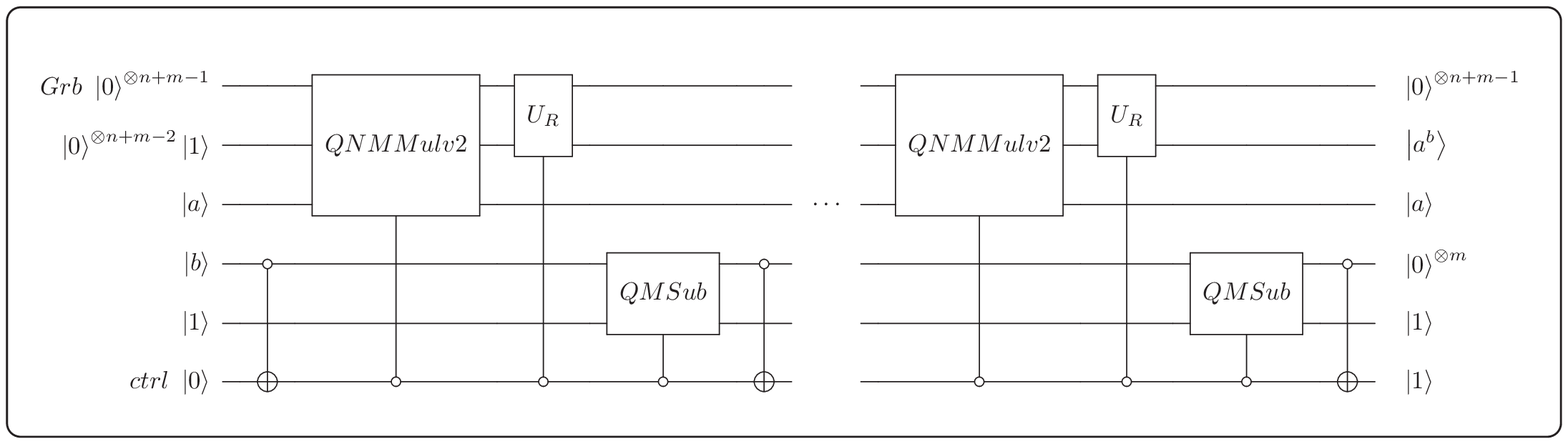}}%
		\hspace{1mm}%
		\subfloat[]{\includegraphics[width=.27\textwidth]{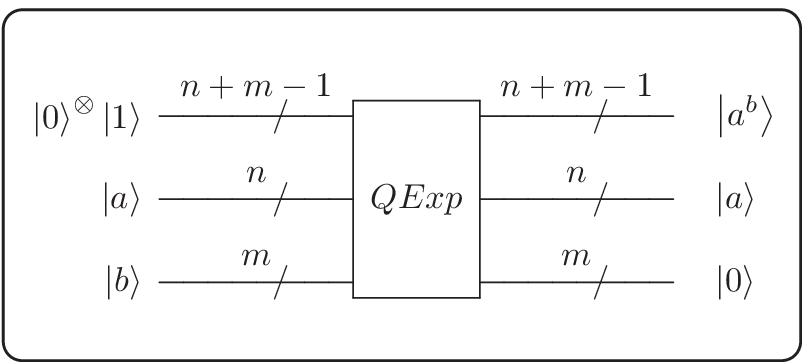}}}%
	\vspace*{8pt}%
	\caption{The QFT based exponentiation operation (QExp) on the original input states (a) the quantum circuit of QExp, (b) the simplified graph of QExp.}%
	\label{Fig:19}%
\end{figure}

We can easily perform the exponentiation operation by using the QNMMulv2 circuit. According to $n$-qubits $\ket{a}$ and $m$-qubits $\ket{b}$, maximum $n+m-1$-qubits are needed for the the result of $\ket{a^b}$. Firstly the states $\ket{0}^{\otimes n+m-1}$ for garbage as $Grb$, $\ket{a}$ and $\ket{0}^{\otimes n+m-2}\ket{1}$ for the result are prepared as inputs of the QNMMulv2 circuit. Then the states $\ket{b}$ and $\ket{1}$ are prepared to count $b$ repetition of the multiplication operation and to control the end of the process. After the multiplication with QNMMulv2, the temporary result is stored in garbage qubits and  $\ket{0}$ in the result qubits. After this process, the circuit $U_R$ shown in Fig.~\ref{Fig:18} is applied and the garbage and the result qubits are swapped for the next operation QNMMulv2. The operation $U_R$ works like the operation SWAP. However, the difference of the operation $U_R$ from the operation SWAP is that; for the swapping of the two qubits, three CNOT gates are used in the operation SWAP and two CNOT gates are used in the operation $U_R$. The operation SWAP swaps any two states, while the proposed operation $U_R$ swaps only the state $\ket{0}$ and any another state. In other words, for the states where one of the inputs is $\ket{0}$, we can say that the operation $U_R$ is a special version of the operation SWAP that uses less resources. In order to repeat the multiplication process $b$ times for the operation $a^b$, the state $\ket{1}$ is subtracted from the state $\ket{b}$ after each multiplication operation and therefore the states $\ket{b}$ and $\ket{1}$ are prepared. Since there is no possibility of overflow in this subtraction operation, the operation QMSub is applied to subtract $\ket{1}$ from $\ket{b}$. A zero-controlled NOT gate is applied to the ancillary qubit $\ket{0}$ as $ctrl$ to check if the state $\ket{b}=\ket{0}^{\otimes m}$. The operations repeat until the state $ctrl$ becomes $\ket{1}$ and the exponentiation operation is completed when the state $ctrl=\ket{1}$. The quantum circuit of QFT-based exponentiation operation (QExp) on two integers ($n \times m$) is shown in Fig.~\ref{Fig:19}.

The proposed QFT-based exponentiation operation requires $n+m-1$-qubits for the result $\ket{a^b}$, $n+m-1$-ancillary qubits for the garbage as $Grb$ and 2-ancillary qubits for $\ket{1}$ (to decrease the state $\ket{b}$) and $\ket{0}$ as $ctrl$. The proposed exponentiation operation operates directly on the original input states and the state $\ket{a}$ remains unchanged from inputs while the state $\ket{b}$ changes to $\ket{0}$. If the state $\ket{b}$ is not desired to be changed, the operations are performed on the copy of the state $\ket{b}$, as in the other proposed operations. Then an additional $m$-ancillary qubits will be needed for the copy of $\ket{b}$. The number of operations required for the $U_R$ circuit is $(2n)$ for $n$-qubits input and the time complexity is $O(n)$. The number of operations required for the $QExp$ circuit is $\left(21n^2+15n+3\left(m^2+m\right)/2+9\right) = \left(12n^2+9n+9\right)$ (for $n=m$) for the inputs $n$-qubits $\ket{a}$ and $m$-qubits $\ket{b}$ and the time complexity is $O(n^2)$. 

\section{Discussion}
\label{sec:4}

There are no non-modular operations on the signed integers ($n \times m$) without any limitation in the literature. In the all studies in the literature, addition, multiplication and exponential operations are performed on integers having the same number of qubits ($n \times n$). There is no study on quantum subtraction, quantum division and quantum two's complement methods. Only the addition operation in Ref.~\citen{Perez2017} (limited for negative integers) and the multiplication operation in Ref.~\citen{Sanchez2008} work with signed integers. All the studies in the literature are modular, except for the addition operation in Ref.~\citen{Perez2017}. Too many $2n$-CNOT gates are used in non-QFT based quantum arithmetic operations. Therefore, the number of basic gates to be used increases too much. Since there are no studies that do exactly the same procedures as the methods proposed in this study, some comparisons with the studies in Refs.~\citen{Perez2017,Sanchez2008}, two nearest studies, will be shown in the tables.

The required numbers of ancillary qubits (on the states $\ket{a}, \ket{b}$ and the copied states $\ket{a^{'}}, \ket{b^{'}}$) of the proposed arithmetic operations are shown in Table~\ref{Tab:1} (the states $\ket{a}$ and $\ket{a^{'}}$ are $n$-qubits, the states $\ket{b}$ and $\ket{b^{'}}$ are $m$-qubits). The required numbers of operations and the time complexities of the proposed arithmetic operations are shown in Table~\ref{Tab:2}. The required number of ancillary qubits, the required numbers of operations and the time complexities of the other proposed operations are shown in Table~\ref{Tab:3}.

\begin{table}[ph]
	\tbl{The required number of ancillary qubits of the proposed modular and non-modular arithmetic operations.\label{Tab:1}}
	{\begin{tabular}{@{}ccc@{}} \toprule
			Operation & On $\ket{a}$ and $\ket{b}$ &On $\ket{a^{'}}$ and $\ket{b^{'}}$ \\
			& ancillary qubits& ancillary qubits\\ 
			\colrule
			QNMAdd &$1$ &$n+1$\\		
			QMAdd &$0$ &$n$\\	
			QNMSub &$1$ &$n+1$\\	
			QMSub &$0$ &$n$\\	
			QNMMul &$n+m+2$ &$n+2m+2$\\	
			QNMDiv &$n+3$ &$2n+m+3$\\
			QExp &$2(n+m)$ &$2n+3m$\\			
			\botrule
	\end{tabular}}
\end{table}

\begin{table}[ph]
	\tbl{The required numbers of operations and the time complexities of the proposed modular and non-modular arithmetic operations.\label{Tab:2}}
	{\begin{tabular}{@{}ccc@{}} \toprule
			Operation &numbers of &time \\
			&operations &complexity \\
			\colrule
			QNMAdd &$\left(n^2+3n+18+\frac{1}{2}(m(2n-m+3))\right)$&$O\left(n^2\right)$\\		
			QMAdd &$\left(n^2+n+\frac{1}{2}(m(2n-m+1))\right)$&$O\left(n^2\right)$\\	
			QNMSub &$\left(n^2+3n+18+\frac{1}{2}(m(2n-m+3))\right)$&$O\left(n^2\right)$\\	
			QMSub &$\left(n^2+n+\frac{1}{2}(m(2n-m+1))\right)$&$O\left(n^2\right)$\\	
			QNMMul &$\left(\frac{1}{2}(5n^2+n)+4m^2+4nm+6m+7\right)$&$O\left(n^2\right)$\\
			QNMDiv &$\left(3n^2+12n+nm+\frac{1}{2}(m^2+11m)+36\right)$&$O\left(n^2\right)$\\
			QExp &$\left(\frac{1}{2}\left(21n^2+15n+3\left(m^2+m\right)\right)+9\right)$&$O\left(n^2\right)$\\
			\botrule
	\end{tabular}}
\end{table}

\begin{table}[ph]
	\tbl{The required numbers of ancillary qubits, required numbers of operations and the time complexities of the other proposed operations.\label{Tab:3}}
	{\begin{tabular}{@{}cccc@{}} \toprule
			Operation &ancillary &numbers of &time \\
			&qubits &operations &complexity \\
			\colrule
			QTC & $1$ &$\left(n^2+3n\right)$&$O\left(n^2\right)$\\
			QABS & $2$ &$\left(n^2+3n+2\right)$&$O\left(n^2\right)$\\
			QComp & $3$ &$\left(n^2+3n+41+\frac{1}{2}(m(2n-m+3))\right)$&$O\left(n^2\right)$\\
			\botrule
	\end{tabular}}
\end{table}

All of the available studies do the arithmetic operations with integers having the same qubit number ($n \times n$). Therefore, the situations of the proposed circuits on integers of the same qubit numbers were considered for a more efficient comparison. The comparison of the QFT-based proposed addition and the QFT-based addition operation in Ref.~\citen{Perez2017} in terms of the required number of ancillary qubits and operations, the time complexity for signed integers ($n \times n$) is shown in Table~\ref{Tab:4}.

\begin{table}[ph]
	\tbl{The required number of qubits and operaions, the time complexity for addition operation with signed integers ($n \times n$).\label{Tab:4}}
	{\begin{tabular}{@{}cccc@{}} \toprule
			Operation &ancillary &numbers of &time \\ 
			&qubits &operations &complexity \\
			\colrule
			Proposed & $1$ & $\left(\frac{3}{2}(n^2+3n+12)\right)$&$O\left(n^2\right)$\\
			Ref.~\citen{Perez2017} & $1$ & $\left(\frac{3}{2}(n^2+3n+2)\right)$&$O\left(n^2\right)$\\
			\botrule
	\end{tabular}}
\end{table}

The proposed and the existing in Ref.~\citen{Perez2017} QFT-based addition operations require the same number of ancillary qubits. The existing addition operation in Ref.~\citen{Perez2017} uses only 15 gates less than the proposed used. However, this is for the maximum number of qubits of integer $b$. As the number of qubits of integer $b$ decreases, the our proposed addition operation uses fewer resources than Ref.~\citen{Perez2017}. Furthermore, the our proposed addition operation can operate without any limitation on negative integers. 

The comparison of the proposed multiplication and the multiplication operation in Ref.~\citen{Sanchez2008} in terms of the required numbers of ancillary qubits and operations, the time complexity for signed integers ($n \times n$) is shown in Table~\ref{Tab:5}.

\begin{table}[ph]
	\tbl{The required number of ancillary qubits and operations,  the time complexity for multiplication operation with signed integers ($n \times n$).\label{Tab:5}}
	{\begin{tabular}{@{}cccc@{}} \toprule
			Operation &ancillary &numbers of &time  \\ 
			&qubits &operations &complexity \\
			\colrule
			Proposed & $2n+2$ & $O\left(\frac{1}{2}(21n^2+13n)+7\right)$&$O\left(n^2\right)$\\
			Existing~\citen{Sanchez2008} & $n^2+4n$ &$O\left(11n^2+6n+4\right)$&$O\left(n^2\right)$\\
			\botrule
	\end{tabular}}
\end{table}

The proposed QFT-based multiplication operation requires far less ancillary qubits than the existing multiplication based on the classic Booth algorithm in Ref.~\citen{Sanchez2008} for the result. In Ref.~\citen{Sanchez2008}, as the number of qubits of the inputs $\ket{a}, \ket{b}$ increases, the number of required ancillary qubits will be increase exponentially. In terms of time complexity, multiplication operations are almost identical. Considering the non-modular multiplication of the signed numbers and the number of required ancillary qubits, it is seen that the proposed study is more efficient than Ref.~\citen{Sanchez2008}.

The comparison of the proposed comparison and the existing comparison operation in Ref.~\citen{Wang2012} in terms of the required number of ancillary qubits and operations, the time complexity for integers ($n \times n$) is shown in Table~\ref{Tab:6}.

\begin{table}[ph]
	\tbl{The required number of ancillary qubits and operations, the time complexity for comparison operation with integers ($n \times n$).\label{Tab:6}}
	{\begin{tabular}{@{}cccc@{}} \toprule
			Operation &ancillary &numbers of &time \\ 
			&qubits &operations &complexity \\
			\colrule
			Proposed & $3$ & $\left(\frac{1}{2}(3n^2+9n+82)\right)$&$O\left(n^2\right)$\\
			Ref.~\citen{Wang2012} & $2n$ &$\left(48n^2-48n+16\right)$&$O\left(n^2\right)$\\
			\botrule
	\end{tabular}}
\end{table}

The circuit in Ref.~\citen{Wang2012} is designed to make comparisons only with unsigned integers. The proposed circuit is designed to compare all signed integers. While the proposed circuit requires only 3 ancillary qubits for the result, the circuit in Ref.~\citen{Wang2012} requires $2n$ ancillary qubits. The proposed circuit is also better than Ref.~\citen{Wang2012} in terms of time complexity. 

The most important difference of the proposed circuits from the other circuits is that they perform non-modular operations on all signed integers by using less ancillary qubits and lower time complexity. 

\section{Conclusion}
\label{sec:5}

Arithmetic operations are required for most methods using quantum computation, especially quantum image and audio processing. This paper presents quantum circuits of QFT-based non-modular addition, subtraction, multiplication,division and exponentiation operations for signed integers with different qubit numbers. In addition, quantum circuits of two's complement, absolute value and comparison operations are also presented. The proposed procedures in this paper will reduce the resource usage of most quantum studies in the literature. It is also considered that the inputs do not change in order to be useful in many quantum methods to be developed. After this paper, it is aimed to develop quantum edge detection, steganography and pattern recognition algorithms in quantum images by using these quantum arithmetic operations.

\section*{Acknowledgments}
I would like to thank referees for valuable suggestions.

\end{document}